\documentclass[prd,nofootinbib,showpacs,showkeys,preprintnumbers,twocolumn]{revtex4}

\usepackage{graphicx}
\usepackage{amsmath}
\usepackage{amssymb}
\usepackage{amsfonts}
\usepackage{latexsym}
\usepackage{mathrsfs}
\usepackage{color}
\usepackage{slashed}
\usepackage{float}
\usepackage{feynmf} 
\usepackage{feyn}
\usepackage[all]{xy}
\usepackage{hyperref}
\usepackage{dsfont}

\providecommand{\eqn}{Eq.\,}
\providecommand{\eqns}{Eqs.\,}
\providecommand{\fig}{Fig.\,}

\providecommand{\sect}{Sec.\,}
\providecommand{\app}{Appendix\,}
\providecommand{\tabl}{Table\,}

\global\long\def\m{\mu}
\global\long\def\e{\epsilon}
\global\long\def\n{\nu}

\global\long\def\l{\lambda}
\global\long\def\v{\delta}
\global\long\def\G{\Gamma}
\global\long\def\r{\rho}
\global\long\def\a{\alpha}
\global\long\def\t{\tau}
\global\long\def\b{\beta}
\global\long\def\a{\alpha}
\global\long\def\g{\gamma}
\global\long\def\b{\beta}

\renewcommand{\Re}{\mbox{Re}}

\newcommand{\tr}{\mbox{tr}}

\begin{document}

\preprint{MPP-2013-28}

\title{Improving Electro-Weak Fits with TeV-scale Sterile Neutrinos}

\author{E. Akhmedov$^{a}$}
\email[\,]{evgeny.akhmedov@mpi-hd.mpg.de}    

\author{A. Kartavtsev$^{b}$}
\email[\,]{alexander.kartavtsev@mpp.mpg.de}

\author{M. Lindner$^{a}$}
\email[\,]{manfred.lindner@mpi-hd.mpg.de}

\author{L. Michaels$^{a}$}
\email[\,]{lisa.michaels@mpi-hd.mpg.de}    

\author{J. Smirnov$^{a}$}
\email[\,]{juri.smirnov@mpi-hd.mpg.de}

\affiliation{%
$^{a}$Max-Planck-Institut f\"ur Kernphysik, Saupfercheckweg 1, 69117 Heidelberg, Germany\\
$^{b}$Max-Planck-Institut f\"ur Physik, F\"ohringer Ring 6, 80805 M\"unchen, Germany}

\begin{abstract}
We study the impact of TeV-scale sterile neutrinos on electro-weak  precision observables and lepton  number and flavour violating decays
in the framework of a type-I see-saw extension of the Standard Model. At tree level sterile neutrinos manifest themselves 
via non-unitarity of the PMNS matrix and at one-loop level they modify the oblique radiative corrections. We 
derive explicit formulae for the $S,T,U$ parameters in terms of the neutrino masses and mixings and 
perform a numerical fit to the electro-weak observables. We find regions of parameter space with a sizable active-sterile 
mixing which provide a better over-all fit compared to the case where the mixing is negligible. Specifically we find improvements of the invisible $Z$-decay width, 
the charged-to-neutral-current ratio for neutrino scattering experiments and of the deviation of the $W$ boson mass from the theoretical 
expectation. 
\end{abstract}

\maketitle

\section{\label{Introduction}Introduction}

The Standard Model (SM) is extremely successful and has passed numerous experimental tests. 
Moreover the last missing piece, the Higgs particle, has recently likely been seen by 
the ATLAS and CMS collaborations \cite{CMS:2012gu,ATLAS:2012gk}. On the other hand,  the SM 
is for a number of theoretical reasons incomplete. It also does not explain three known 
experimental facts, namely the tiny active neutrino masses, the baryon asymmetry of the 
Universe and the existence of dark matter. 
A simple yet elegant way to solve two or even all of these problems is to supplement 
the SM by three singlets:
\begin{align} 
	\label{lagrangian}
	\mathscr{L}=\mathscr{L}_{SM}&+
	{\textstyle\frac12}\bar N_i \bigl(i\slashed{\partial} - M_i\bigr)N_i\nonumber\\
	&- h_{\alpha i}\bar \ell_\alpha {\tilde \phi}  N_i
	-h^\dagger_{i\alpha} \bar N_i {\tilde \phi}^\dagger  \ell_\alpha\,,
\end{align}
where $N_i=N^c_i$ are Majorana fields, $\ell_\alpha$ are the lepton doublets, 
$\tilde \phi\equiv i\sigma_2 \phi^*$ is the conjugate of the Higgs doublet, and $h$ are 
the corresponding Yukawa couplings. 

This is a small modification of the SM, also because so far the principles which select the 
fermionic representations in a given theory are (besides for anomaly conditions) unknown. 
Nevertheless, this modification allows to accommodate those three experimental facts:
First, after breaking of the electro-weak symmetry the active neutrinos acquire masses via 
the type-I see-saw mechanism \cite{Minkowski:1977sc,Yanagida:1979as,Mohapatra:1979ia,
GellMann:1980vs}. Second, within the same setup the observed  baryon asymmetry of the 
Universe can be naturally explained by leptogenesis \cite{Fukugita:1986hr}. According to this 
scenario the lepton asymmetry which is produced by the heavy neutrinos is converted into 
the baryon asymmetry by anomalous electro-weak processes \cite{'tHooft:1976up,
Klinkhamer:1984di,Kuzmin:1985mm}. Depending on the values of the masses and couplings 
in \eqref{lagrangian} the generation of the asymmetry proceeds either through  decays 
\cite{Buchmuller:2004nz,Giudice:2003jh,Davidson:pr2008,Blanchet:2012bk,Garny:2009rv,
Garny:2009qn,Garny:2010nj,Beneke:2010wd,Garny:2010nz,Garbrecht:2010sz,Drewes:2012ma,
Frossard:2012pc,Dick:1999je} or via oscillations of the heavy neutrinos \cite{Akhmedov:1998qx}. 
The extension of the SM spectrum by sterile neutrino states may even provide a 
solution to the third problem, since sterile neutrinos are a perfect dark matter 
candidate (see e.g. \cite{Shi:1998km,Asaka:2005pn,Shaposhnikov:2008pf,Boyarsky:2009ix,
Bezrukov:2009th,Bezrukov:2012as,Canetti:2012vf,Canetti:2012kh}). A further possible implication of sterile neutrinos is a modified active neutrino flux in reactor experiments \cite{Lasserre:2012jc}. This could explain the observed reduced electron antineutrino fluxes \cite{Kopp:2011qd}.

Note that the addition of singlets to the SM has actually important consequences.
Not only are there many new parameters, but also the global symmetries are changed,
since lepton number is broken. Furthermore the single scale of the SM, namely the electro-weak 
VEV, is amended by the new mass scales $M_i$. Within the type-I see-saw the sterile neutrinos
pick up masses proportional to $M_i$ and these could in principle have any value which is not 
excluded by experiment. It is therefore interesting to study implications of sterile neutrinos 
assuming any mass. Some effects exist even for ultra heavy sterile neutrinos, while some are
only phenomenologically important for not so heavy states, which are in the mass range of several TeV or lower.
 
The existence of Majorana neutrinos has well known consequences on the phenomenology 
below the electro-weak scale. In particular, the new states can contribute to the amplitude of the neutrinoless
double-beta decay \cite{Rodejohann:2012xd,LopezPavon:2012zg,Mitra:2012qz,Mitra:2011qr,
Blennow:2010th} and induce rare charged lepton decays \cite{Antusch:2006vwa,Dinh:2012bp}.
Furthermore, they can affect the electro-weak precision observables  (EWPOs) via tree-level as well 
as loop contributions and thus provide an explanation for anomalies in the experimental data. 
In particular, the tree-level effects result in   non-unitarity of the active neutrino mixing matrix 
\cite{Antusch:2006vwa} and lead to a suppression of the invisible $Z$-decay width. 
This is in agreement with the long standing fact that the LEP measurement of the invisible $Z$-decay 
width is two sigma below the value expected in the SM \cite{Beringer:1900zz}. Furthermore 
the neutral-to-charged-current ratio in neutrino scattering experiments can be changed thus providing an explanation for the 
NuTeV anomaly \cite{NuTeV}. Also a slight shift of the $W$ boson mass from the
value derived from other SM parameters is induced, reducing the tension between the input 
parameters of the electro-weak fit and the experimentally observed value \cite{Ferroglia:2012ir}. 
These electro-weak observables are affected not only at tree-level but also by loop effects. The latter can be taken into account in the form of the oblique corrections and can partially `screen' the tree-level contributions, as we will see later. The impact of oblique corrections on models with non-universal neutrino gauge coupling has been studied in \cite{Takeuchi:A,Takeuchi:B,Takeuchi:2004dq}. In these works the tree-level effects
($\e_e$, $\e_\mu$ and $\e_\tau$) and oblique ($S$, $T$ and $U$) corrections have been treated 
as independent parameters. However, in the model described by the Lagrangian \eqref{lagrangian} 
they are functions of the Majorana masses and Yukawa couplings and are therefore not independent. 

Encouraged by the fact that sterile neutrinos are very well motivated we study their phenomenological impact in this paper. Specifically we consider TeV-scale sterile neutrinos with a sizable active-sterile mixing and   determine their over-all contributions to the EWPOs and to indirect detection experiments
in the framework of the see-saw type-I extension of the SM. In contrast to the previous
studies we derive expressions for the $S,T,U$-parameters in terms of the masses and mixing 
angles of the Majorana neutrinos, thus all quantities effecting the observables are functions of the parameters in \eqref{lagrangian}. Our approach is therefore completely self-consistent. In \sect\ref{LowEnergyObservables} we discuss the influence of heavy Majorana neutrinos on the phenomenology below the electro-weak 
scale and   provide expressions for the $S,T,U$-parameters. Next we perform in 
\sect\ref{Fit} a likelihood fit to the EWPOs taking into account up-to-date experimental 
constraints including neutrino oscillation data and limits from the non-observation of rare 
charged lepton decays and neutrinoless double-beta decay. We find that TeV-scale sterile neutrinos 
improve the overall fit by bringing the invisible $Z$-decay width, the charged-to-neutral 
current ratio  for neutrino scattering  and the $W$ boson mass in agreement with the experimentally observed values within the experimental precision. 
The best-fit regions provide testable experimental signatures. For the normal- and quasi-degenerate light neutrino mass spectra we find that $0\nu\beta\beta$ decay rates are close to the current experimental sensitivities.
For the inverted hierarchy the mass region is such that part of the parameter space can be tested at the LHC after the 14 TeV upgrade.
Finally, in \sect\ref{Summary} we summarize the obtained results and present our conclusions.

\section{\label{LowEnergyObservables} Observables}

The heavy neutrino fields in \eqref{lagrangian} are SM singlets and do not participate in the gauge interactions. 
However, the breaking of the electroweak symmetry induces a neutrino Dirac mass term. The mass eigenstates
obtained by diagonalization of the full mass matrix couple to the $Z$ and $W$ bosons. Expressed in terms of the mass eigenstates the corresponding part of the Lagrangian takes the form 
\begin{align}
\label{gaugelagrangian}
&\mathscr{L}_\text{int}=-\frac{e}{2c_ws_w}Z_\mu{\textstyle\sum_{i,j=1}^{3+n}}{\textstyle\sum}_{\alpha=e,\m,\t}
\bar{\nu}_i \mathbf{U}^\dagger_{i\alpha}\gamma^\mu P_L \mathbf{U}_{\alpha j} \nu_j\nonumber\\
&-\frac{e}{\sqrt{2}s_w} W_\mu{\textstyle\sum_{i=1}^{3+n}}{\textstyle\sum}_{\alpha=e,\m,\t}\bar{\nu}_i 
\mathbf{U}^\dagger_{i\alpha}\gamma^\mu P_L e_\alpha + {\rm h.c.}\,,
\end{align}
where $e_\alpha$ denote the charged leptons, $\nu_i$ denote  the light (for $i\leq 3$) 
as well as heavy (for $4 \leq  i \leq 3+n$) 
neutrinos, and $\mathbf{U}$ is the full unitary $(3+n)\times (3+n)$ neutrino mixing matrix. 
To give masses to at least two active neutrinos, as required by oscillation 
experiments, we need two or more heavy neutrinos, i.e. $n\geq 2$, but otherwise $n$ 
is unconstrained. Below we review the phenomenology of 
\eqref{gaugelagrangian} together with the corresponding up-to-date experimental results. 

\textit{Lepton-flavor violating decays.}
In the scenario under consideration the branching ratio of $\mu\rightarrow e\gamma$  
decay is gi\-ven by \cite{ChengLi} 
\begin{align}
& \text{BR}(\mu\rightarrow e\gamma)=
\frac{\Gamma(\mu\rightarrow e\gamma)}{\Gamma(\mu\rightarrow e\nu\bar{\nu})}=
\frac{3\alpha}{32\pi}|\delta_{\nu}|^{2}\,,
\end{align}
 where $\delta_{\nu}=2\sum_{i}\mathbf{U}_{ei}^{\ast}\mathbf{U}_{\mu i}\,g\left(m_{i}^{2}/M_{W}^{2}\right)$
and the loop function $g$ is defined by
\begin{align*}
& g(x)=\int_{0}^{1}\frac{(1-\alpha)d\alpha}{(1-\alpha)+\alpha x}[2(1-\alpha)(2-\alpha)+\alpha(1+\alpha)x].
\end{align*}
Note that we use $m_i$ to denote the masses of both light and heavy neutrinos.
Since the masses of the active neutrinos are very small we can neglect 
them in the loop integral, $g(m_i^2/M_W^2)\approx g(0)= 5/3$. Using   unitarity of 
the full mixing matrix we then find
\begin{align}
\label{delta_full}
\delta_{\nu}=2\, \textstyle{\sum_{i=4}^{3+n}} \mathbf{U}_{e i}^{\ast}\mathbf{U}_{\mu i}
\left[g\left(m_{i}^{2}/M_{W}^{2}\right)-5/3\right]\,. 
\end{align}
The recent  limit on this branching ratio obtained by the MEG collaboration \cite{MEG} is
\begin{align}
\label{delta_exp}
\text{BR}(\mu^{+}\rightarrow e^{+}\gamma)\leq2.4\cdot10^{-12}\,
\end{align}
at $90\%$ confidence level. From \eqref{delta_full} and \eqref{delta_exp} we can infer 
bounds  on the products of the mixing elements $\mathbf{U}_{\mu i}$ and $\mathbf{U}_{e i}$. 
An analogous relation also exists for the $\tau\rightarrow e\gamma$ decay. However, the 
corresponding experimental constraints are much weaker and will not be considered 
here.

\textit{Neutrinoless double-beta decay.}
Neutrinoless double-beta decay constrains the effective mass of the electron neutrino $\langle m_{ee}\rangle$. 
The latter receives contributions from the light as well as from the heavy mass eigenstates 
\cite{Rodejohann:2012xd}:
\begin{align}
|\langle m_{ee}\rangle|\approx\bigl|
\textstyle{\sum_{i=1}^3}\mathbf{U}_{ei}^{2}m_i-
\textstyle{\sum_{i=4}^{3+n}} F(A,M_i)\mathbf{U}_{ei}^{2}m_i\bigr|\,.
\end{align}
For masses of the heavy neutrinos in the TeV range one can use an approximation: 
$F(A,m_i)\approx (m_a/m_i)^2 f(A)$, where $m_a\approx 0.9$ GeV and $f(A)$ depends on the 
decaying isotope under consideration \cite{Blennow:2010th,Ibarra:DoubleBeta}. A conservative 
bound, $|\langle m_{ee}\rangle|<0.4$  eV, has been recently obtained by EXO collaboration 
\cite{Auger:Beta}. 

\textit{Unitarity and lepton universality violation.} For the following analysis it is convenient 
to represent $\mathbf{U}$ in the form
\begin{equation}
\label{mixingmatrix}
\mathbf{U}=\left(
\begin{array}{cc}
\mathscr{U} & \mathscr{R}\\
\mathscr{W} & \mathscr{V}
\end{array}
\right)\,.
\end{equation}
The $(n\times 3)$ matrix $\mathscr{R}$ describes the active-sterile mixing. An obvious 
consequence of a nonzero active-sterile mixing is that the $(3\times3)$ PMNS 
matrix $\mathscr{U}$ is no longer exactly unitary \cite{Antusch:2006vwa}. The deviation from 
unitarity can be parameterized by
\begin{align}
\label{epsilon}
\e_\alpha\equiv {\textstyle\sum_{i\geq 4}} |\mathbf{U}_{\alpha i}|^2\,.
\end{align}
In general the quantities $\e_e$, $\e_\mu$ and $\e_\tau$ are not equal. In other words, 
there is also a violation of lepton universality. Experimental bounds on linear combinations 
of the $\epsilon_\alpha$ read \cite{Takeuchi:A}
\begin{subequations}
\label{NonUniversConstr}
\begin{align}
\e_e-\e_\m=0.0022 \pm 0.0025\,, \\
\e_\m-\e_\t=0.0017 \pm 0.0038\,,\\
\e_e-\e_\t=0.0039 \pm 0.0040\,.
\end{align}
\end{subequations}
The stringent experimental bound on the $\mu\rightarrow e\gamma$ branching 
ratio implies that either $\e_e$ or $\e_\m$ is negligible in \eqref{NonUniversConstr}. For a recent study of lepton universality violation in Pion and Kaon decays in the presence of sterile neutrinos consult \cite{Abada:2012mc}. 
 
\textit{Heavy neutrinos at colliders.} Heavy singlet Majorana fermions have been searched
for at the LEP and LHC colliders, see \cite{TaoHan, Perez:2009mu} for a review. The searches were based on
the production of the heavy state due to a considerable mixing to the active neutrinos.

At LEP the heavy neutrino could be produced in $e^+e^-$ annihilation, $e^+e^-\rightarrow N \nu$,
via $s$-channel $Z$-exchange as well as via $t$-channel $W$-exchange. The produced Majorana neutrinos
then rapidly decay via the weak neutral or charged currents: $N\rightarrow Z\nu$ and $N\rightarrow W e$. 
A search for heavy neutrinos with masses up to $\sim 200$ GeV has been performed by the L3 and Delphi collaborations
using the latter decay channel with $W$ decaying into hadrons \cite{LEP, HeavyBound}. The
experimental signature of these events would be one isolated electron plus hadronic jets.

\begin{figure}[h]
\centering
\includegraphics[width=0.7 \columnwidth]{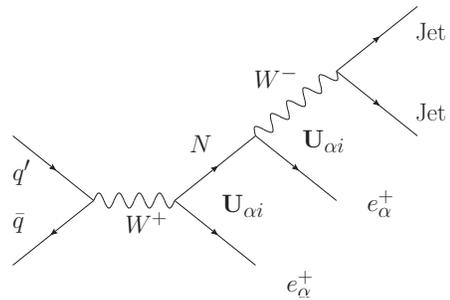}
\caption{\label{DirectNProcess}  Lepton number violating process mediated by the sterile neutrino, see e.g. \cite{Keung:1983uu}.}
\end{figure}

New limits on the active-sterile mixing for the heavy neutrino masses up to 210 GeV have been
obtained recently by the CMS collaboration \cite{CMS:2012pz} using a dilepton decay channel with
two leptons of equal charge and flavour plus jets, see \fig\ref{DirectNProcess}.
Violation of lepton number in this process occurs due to the Majorana nature of the sterile
neutrino. For large Majorana masses the square of the momentum transfer in the propagator of the
intermediate neutrino can be neglected and the production cross section depends on the combination
$|\sum_i \mathbf{U}^2_{\alpha i}\,m^{-1}_i|$. Assuming that only one of the heavy neutrinos couples
to the electron with a strength $|\mathbf{U}_{ei}|^2\approx 5.2\cdot 10^{-3}$ it has been estimated
in \cite{Almeida:2000pz} that after the LHC upgrade to $\sqrt{s} \approx 14$ TeV this channel can be
used to search for Majorana neutrinos with masses up to roughly 800 GeV. For these parameters we get
$|\sum_i \mathbf{U}^2_{\alpha i}\,m^{-1}_i|\simeq 6.5\cdot 10^{-3}$ TeV$^{-1}$ and one
event per $100\, \text{fb}^{-1}$ is expected. Even though this is a very weak statistical signal
it has been noted that in the highest mass region $(800\, \text{GeV})$ the leptons are emitted back to back providing
a very clean signature and allowing for an excellent background suppression. If the luminocity
of $\sim 400\, \text{fb}^{-1}$ is reached one can hope to find the heavy neutrinos even if
the above defined combination of the mixings and masses is as small as $3.25\cdot 10^{-3}$ TeV$^{-1}$.

A possible way of testing even higher mass ranges is to study electron-proton collisions where the dominant
detection channel is $e^- + q \rightarrow q' + e^+ + W$ with the Majorana neutrino as an intermediate state.
In \cite{deAlmeida:2002pr} it has been estimated that at an electron-positron collider with a
center of mass energy $\sqrt{s} \approx 6$ TeV mass ranges up to 1.5 TeV can be tested.

\textit{Non-unitarity in neutrino oscillations.}
The standard oscillation formula assumes unitarity of the PMNS matrix $\mathscr{U}$, which 
is violated in the model under consideration. The non-unitarity of $\mathscr{U}$ will manifest 
in a modified oscillation probability formula \cite{Antusch:2006vwa}, which in particular 
includes a zero-length effect \cite{Langacker:1988up}:
\begin{equation}
\label{Zerolength}
P_{\alpha \beta}(L=0) = 
\frac{\delta_{\alpha \beta} \left( 1 - 2 \e_\alpha \right) + \e_\alpha \e_\beta}{(1-\e_\alpha )(1-\e_\beta)}\,.
\end{equation}
Being of second order in $\e$ for $\alpha \neq \beta$ this effect is strongly suppressed. As discussed by a number of authors, see \cite{Antusch:2006vwa} and references therein, the probabilities do not add up to unity. 
At non-zero distances the non-unitarity effects are linear in $\e$ \cite{Antusch:2009pm} 
and may affect the results of global fits. For the purposes of this work we will use the best-fit results for the light 
neutrino parameters obtained in a unitary fit as an input, see \cite{Thomas}. 
The order $\e$ corrections to the parameters of the active neutrinos will be computed using 
\eqref{active-sterile-mixing}. At the end, when comparing the corrected values to the results of the 
 unitary fit in \cite{Thomas} we find agreement within the one sigma intervals. The values are also consistent with the one sigma intervals obtained in a non-unitary fit \cite{Antusch:2006vwa}. Furthermore, we have checked that our results are not sensitive to changes  of the active neutrino parameters within their 
experimental errors, which justifies our procedure.

\textit{Electroweak precision observables.}
An important consequence of non-unitarity is that the couplings of the
light neutrinos to the $Z$ and $W$ bosons are suppressed with respect 
to their SM values. Taking this effect into account we find that the 
invisible $Z$-decay width is suppressed:
\begin{equation}
\label{GInvTreeLevel}
\Gamma_\text{inv}/\left[\Gamma_\text{inv}\right]_{\text{SM}}=
{\textstyle\frac13\sum_\alpha}(1-\e_\alpha)^2\,. 
\end{equation}
This means that the `effective number of neutrinos' measured in the $Z$-decay is slightly 
less than three, which is qualitatively agreeing with the LEP results \cite{Beringer:1900zz}.
\begin{table}[h]
\begin{tabular}{l|l|l}
\hline 
EWPO & Theory (Standard Model) & Experiment\\
\hline \hline 
$\Gamma_{\rm lept}$ (MeV) & $84.005\pm0.015$ & $83.984\pm0.086$\\
$\Gamma_{\rm inv}/\Gamma_{\rm lept}$ & $5.9721\pm0.0002$ & $5.942\pm0.016$\\
$\sin^2\theta_{\text{W}}$ & $0.23150\pm0.0001$ & $0.2324\pm0.0012$\\
$g_{\text{L}}^{2}$ & $0.3040\pm0.0002$ & $0.3026\pm0.0012$\\
$g_{\text{R}}^{2}$ & $0.0300\pm0.0002$ & $0.0303\pm0.0010$\\
$M_{\text{W}}$ (GeV) & $80.359\pm0.011$ & $80.385\pm0.015$\\
\hline 
\end{tabular}
\caption{\label{ElectrowekObservables} Theoretical predictions and experimental 
results for electro-weak precision observables (EWPO). The theoretical predictions
are taken from \cite{Gfitter}. The experimental values for the invisible and 
leptonic $Z$-decay widths as well as the Weinberg angle are from \cite{Beringer:1900zz}. 
For the Weinberg angle we use the value measured in the hadronic processes to make sure
that it is free of the non-unitarity corrections. The values of $g_{L}$ and $g_{R}$ are 
taken from the NuTeV results after including a recent NNLO analysis \cite{NuTeV, NNLO}.}
\end{table} 
Similarly, cross-sections of the charged- and neutral-current neutrino scatterings 
on quarks are also affected:
\begin{subequations}
\label{CCNCcrossection}
\begin{align}
\label{CCrossection}
{\sigma}^{\text{CC}}_\alpha&={\sigma}^{\text{CC}}_{\alpha, \text{SM}}(1-\e_\alpha)\,,\\
\label{NCrossection}
{\sigma}^{\text{NC}}_\alpha&={\sigma}^{\text{NC}}_{\alpha,\text{SM}}(1-\e_\alpha)^2\,. 
\end{align}
\end{subequations}
The stronger relative suppression of the neutral current interactions  can 
be observed in experiments measuring ratios of the corresponding cross 
sections. In particular, it is qualitatively consistent with the results of the 
NuTeV experiment \cite{Zeller:2001hh,Zeller:2002du,McFarland:2002rk,Zeller:2002et}. 
Another important consequence of non-unitarity is that $G_\mu$ -- the Fermi constant 
measured in the muon decay -- is not equal to the Fermi constant measured in experiments 
with semi-leptonic processes, but
\begin{eqnarray}
\label{GFermi}
G^2_\mu= G^2_F\, (1-\e_\mu) (1-\e_e) \,. 
\end{eqnarray}
Since the muon decay width is used as an input in the SM fits, this modification 
influences many observables and has been used in \cite{Antusch:2006vwa} to obtain 
bounds on the active-sterile mixing. However, as argued in \cite{Takeuchi:A, 
Takeuchi:B}, the impact of the heavy neutrinos is not limited to the above discussed 
tree-level effects. Heavy neutrinos contribute to  the self-energies of 
the $W$ and $Z$ bosons and therefore modify their propagators. 
These loop effects can be described in terms of the $S,T,U$ parameters 
\cite{Peskin:1991sw}. A more detailed description of the oblique correction formalism is presented in \app\ref{DefOfSTU}, while the calculations are shown in \app\ref{STUcalculation}. Combining the tree-level and one-loop contributions one obtains 
\cite{Takeuchi:A, Takeuchi:B}
\begin{subequations}
\label{EWobservables}
\begin{align}
\label{Glept}
\frac{\G_{\text{lept}}}{\left[\G_{\text{lept}}\right]_{\text{SM}}}&=1
+0.6\,(\e_{e}+\e_\m+0.0145\,T)\nonumber\\
&-0.0021\, S\,, \\
\label{Ginv}
\frac{\G_{\text{inv}}/\G_{\text{lept}}}{\left[\G_{\text{inv}}/\G_{\text{lept}}\right]_{\text{SM}}}&=1
-0.67\,(\e_{e}+\e_{\m}+\e_{\t})\nonumber\\
&+0.0021\, S-0.0015\, T \,, \\
\label{sinTw}
\frac{\sin^{2}\theta_{\text{w}}^{\text{lept}}}{\bigl[\sin^{2}\theta_{\text{w}}^{\text{lept}}\bigr]_{\text{SM}}}&=1
-0.72\,(\e_{e}+\e_\m+0.0145\,T) \nonumber\\
&+0.0016\, S\,,\\
\label{gL}
\frac{g_{L}^{2}}{\left[g_{L}^{2}\right]_{\text{SM}}}&=1
+0.41\,\e_{e}-0.59\,\e_{\m}\nonumber\\
&-0.0090\, S+0.0022\, T\,, \\
\label{gR}
\frac{g_{R}^{2}}{\left[g_{R}^{2}\right]_{\text{SM}}}&=1
-1.4\,\e_{e}-2.4\,\e_{\m}\nonumber\\
&+0.031\, S-0.0067\, T\,,\\
\label{Mw}
\frac{M_{\text{W}}}{\left[M_{\text{W}}\right]_{\text{SM}}} &= 1
+ 0.11\,\e_{e}+0.11\,\e_{\m}\nonumber\\
&-0.0036\, S+0.0056\, T+ 0.0042\, U \,.
\end{align}
\end{subequations}
Comparing \eqref{GInvTreeLevel} and \eqref{Ginv} we see that the first line in the 
latter equation is the leading-order expansion of the former one and that the loop 
corrections enter through the $T$ and $S$ parameters. 

\textit{Cancellation mechanism.} Since $S$ and $U$ are related to derivatives 
of the $W$ and $Z$ boson self-energies whereas $T$ is proportional to a difference of
the two, usually $S$ and $U$ are subdominant in comparison to $T$. If  
$S$ is neglected  then \eqref{Glept} and \eqref{sinTw} depend on the same 
combination of tree-level and one-loop corrections, $\e_e+\e_\m+2\alpha_{em} T$,
where $\alpha_{em}$ is the fine structure constant. The reason is that in this approximation 
the shift of these observables is solely due to a shift in $G_\mu$ \cite{Takeuchi:B}. 
It has been argued in \cite{Antusch:2006vwa} that a shift in 
$G_\mu$ has  dramatic effect on the electro-weak observables and thus $\e_e$ 
and $\e_\mu$ are strongly constrained. However, as follows from \eqref{EWobservables}, 
if the tree-level and radiative contributions are of a similar size  the shift induced by 
the tree-level effects can be `screened' by a sizable negative $T$ parameter. Typically, most of the SM 
extensions result in a positive $T$ parameter. Therefore, it was assumed in 
\cite{Takeuchi:B} that the Higgs mass is much larger than the used reference value, 
$m_H=115$ GeV. However, interpreting the recent discovery of the ATLAS and CMS collaborations as the SM Higgs, this option is excluded. 

\textit{Majorana neutrino contribution to the $S,T,U$ parameters.}
Here we argue that the screening can be realized without invoking any new physics beyond that already 
introduced in \eqref{lagrangian}. As demonstrated in \cite{Kniehl:1992ez} 
for a model with a single heavy neutrino, in a certain range of the parameter space 
Majorana neutrinos can generate a negative contribution to the $T$ parameter. In 
the model with $n$ heavy neutrinos we find
\begin{align}
\label{ResultT}
&T_\text{tot}=T_N+T_{\text{SM}}=-\frac{1}{8\pi s^2_w M_W^2}\nonumber\\
&\times \bigl[{\textstyle\sum_{i,j=1}^{3+n}}{\textstyle\sum}_{\alpha\beta}\,
\mathbf{U}^\dagger_{i\alpha}\mathbf{U}_{\alpha j} 
\mathbf{U}^\dagger_{j\beta}\mathbf{U}_{\beta i}\,
Q(0,m^2_i,m^2_j) \nonumber\\
&+ {\textstyle\sum_{i,j=1}^{3+n}}{\textstyle\sum}_{\alpha\beta}\,
\mathbf{U}^\dagger_{i\alpha}\mathbf{U}_{\alpha j}
\mathbf{U}^\dagger_{i\beta}\mathbf{U}_{\beta j}
m_im_jB_0(0,m^2_i,m^2_j)\nonumber\\
&- 2\,{\textstyle\sum_{i=1}^{3+n}}{\textstyle\sum}_{\alpha}
\mathbf{U}^\dagger_{i\alpha}\mathbf{U}_{\alpha i} \,
Q(0,m^2_i,m^2_\alpha)\nonumber\\
&+{\textstyle\sum}_{\alpha} m^2_\alpha B_0 (0,m^2_\alpha,m^2_\alpha)\bigr]\,,
\end{align}
where $m_\alpha$ denote masses of the charged leptons. To shor\-ten the notation we 
have introduced 
\begin{align}
Q(q^2,&m^2_1,m^2_2)\equiv (D-2)B_{22}(q^2,m_1^2,m_2^2)\nonumber\\
&+q^2\bigl[ 
B_1(q^2,m_1^2,m_2^2)+B_{21}(q^2,m_1^2,m_2^2)
\bigr]\,,
\end{align}
where $B_0$, $B_1$, $B_{21}$ and $B_{22}$ are the usual one-loop functions
\cite{Passarino:1978jh}, $D\equiv 4-2\epsilon$  and $\epsilon\rightarrow 0$.
Details of the calculation can be found in \app\ref{STUcalculation}. Note 
that although the loop functions are divergent and contain an arbitrary scale 
$\mu$, their combination \eqref{ResultT} is finite and independent of $\mu$.
This can be shown using the unitarity of the full mixing matrix and the type-I see-saw 
condition $(m_L)_{\alpha \beta} \equiv \sum_{i=1}^{3+n}\mathbf{U}_{\a i}m_{i}
\mathbf{U}_{i\b}^{T}=0$ with $\alpha, \beta \in \{e, \mu, \tau \}$. 
Therefore, to compute \eqref{ResultT} one can simply drop the $\epsilon^{-1}$ terms in the expansion 
of the loop functions and evaluate the remaining integrals numerically or 
analytically. As indicated in \eqref{ResultT}, the T parameter is a sum of the SM and 
new contributions. The SM contribution can be obtained from \eqref{ResultT} 
by setting the active neutrino masses to zero and taking into account that $\mathscr{R}=0$ and  
$\mathscr{U}$ is unitary in the SM:
\begin{align}
\label{TSM}
T_{\text{SM}}=-\frac{1}{8\pi s^2_w M_W^2}  &
\bigl[3\,Q(0,0,0) - 2{\textstyle\sum}_{\alpha}\,Q(0,0,m^2_\alpha) \nonumber\\
&+{\textstyle\sum}_{\alpha} m^2_\alpha B_0 (0,m^2_\alpha,m^2_\alpha)\bigr]\,.
\end{align}
This expression is also finite. Note that the PMNS matrix does not appear in
\eqref{TSM} since the mixing becomes unphysical for massless neutrinos.  
The $S$ parameter reads
\begin{align}
\label{ResultS}
&S_\text{tot}=S_N+S_{\text{SM}}=-\frac{1}{2\pi M_Z^2}\nonumber\\
&\times \bigl[{\textstyle\sum_{i,j=1}^{3+n}}{\textstyle\sum}_{\alpha\beta}\,
\mathbf{U}^\dagger_{i\alpha}\mathbf{U}_{\alpha j} 
\mathbf{U}^\dagger_{j\beta}\mathbf{U}_{\beta i}\,
\Delta Q(M_Z^2,m^2_i,m^2_j) \nonumber\\
&+ {\textstyle\sum_{i,j=1}^{3+n}}{\textstyle\sum}_{\alpha\beta}\,
\mathbf{U}^\dagger_{i\alpha}\mathbf{U}_{\alpha j}
\mathbf{U}^\dagger_{i\beta}\mathbf{U}_{\beta j}
m_im_j\Delta B_0(M_Z^2,m^2_i,m^2_j)\nonumber\\
&+{\textstyle\sum}_{\alpha} m^2_\alpha B_0(0,m^2_\alpha,m^2_\alpha)
+{\textstyle\sum}_{\alpha} Q(M_Z^2,m^2_\alpha,m^2_\alpha)\nonumber\\
&-2\,{\textstyle\sum}_{\alpha} m^2_\alpha B_0(M_Z^2,m^2_\alpha,m^2_\alpha)
\bigr]\,,
\end{align}
where $\Delta Q(q^2,m^2_1,m^2_2)\equiv Q(0,m^2_1,m^2_2)-Q(q^2,m^2_1,m^2_2)$ and 
$\Delta B_0$ is defined in the same way. The SM contribution, $S_{SM}$, is calculated 
analogously to $T_{SM}$. For the $U$ parameter we obtain
\begin{align}
\label{ResultU}
&U_\text{tot}=U_N+U_{\text{SM}}=\frac{1}{2\pi M_Z^2}\nonumber\\
&\times \bigl[{\textstyle\sum_{i,j=1}^{3+n}}{\textstyle\sum}_{\alpha\beta}\,
\mathbf{U}^\dagger_{i\alpha}\mathbf{U}_{\alpha j} 
\mathbf{U}^\dagger_{j\beta}\mathbf{U}_{\beta i}\,
\Delta Q(M_Z^2,m^2_i,m^2_j) \nonumber\\
&+ {\textstyle\sum_{i,j=1}^{3+n}}{\textstyle\sum}_{\alpha\beta}\,
\mathbf{U}^\dagger_{i\alpha}\mathbf{U}_{\alpha j}
\mathbf{U}^\dagger_{i\beta}\mathbf{U}_{\beta j}
m_im_j\Delta B_0(M_Z^2,m^2_i,m^2_j)\nonumber\\
&+{\textstyle\sum}_{\alpha} m^2_\alpha B_0(0,m^2_\alpha,m^2_\alpha)
-{\textstyle\sum}_{\alpha} Q(M_Z^2,m^2_\alpha,m^2_\alpha)\nonumber\\
&-2\,{\textstyle\sum}_{\alpha} m^2_\alpha B_0(M_Z^2,m^2_\alpha,m^2_\alpha)\nonumber\\
&-2(M_Z/M_W)^2{\textstyle\sum}_{\alpha}\mathbf{U}^\dagger_{i\alpha}\mathbf{U}_{\alpha i}
\Delta Q(M_W^2,m^2_i,m^2_\alpha) 
\bigr]\,.
\end{align}
For a single generation \eqns\eqref{ResultT}, \eqref{ResultS} and \eqref{ResultU} reduce to 
the expressions derived in \cite{Kniehl:1992ez}. The observation that the heavy neutrinos 
can `screen' the tree-level contributions by generating a negative $T$ parameter together 
with the explicit formulae for the $S,T,U$ constitutes one of the main results of the 
present work.

\section{\label{Fit}Fit to the observables}
In this work we investigate the impact 
of TeV-scale sterile neutrinos with a sizable active-sterile mixing on the physics below 
the electro-weak scale in the framework of the see-saw type-I extension of the SM.
In the see-saw limit the active-sterile mixing is of order $\hat m^T_D (\hat M_R)^{-1}$. 
On the other hand, the see-saw formula for the mass matrix of the light neutrinos 
reads $\hat m_\n=-\hat m^T_D (\hat M_R)^{-1} \hat m_D$. Comparing the two expressions 
we conclude that for a large active-sterile mixing the smallness of the active neutrino
masses can not be explained by the scale suppression. Instead,  contributions from different 
heavy states have to mutually cancel. To ensure the cancellation more than one sterile neutrino is needed. 
As already mentioned above, two heavy neutrinos are required to ensure that at least two of 
the light neutrinos are massive. We have also seen that the mixing to either the electron or the 
muon neutrino has to be strongly suppressed to be compatible with the current $\m \rightarrow e \gamma$ 
bounds. Combined together, the two conditions imply that the cancellation is only possible 
for $n \geq 3$. We consider $n=3$ in this paper, which arises naturally in 
the framework of many left-right symmetric or GUT models.

For $n=3$ both $\mathscr{U}$ and $\mathscr{R}$ are $(3\times 3)$ matrices which can be written in the form \cite{Casas:Parameter}:
\begin{subequations}
\label{active-sterile-mixing}
\begin{align}
\mathscr{R}&=-i\,{\cal U}\,\hat{m}_{\rm light}^{\frac12}\,O^{*}\hat{m}_{\rm heavy}^{-\frac12} \, ,\\
\mathscr{U}&=\left(1-\mathscr{R}\, \mathscr{R}^{\dagger}\right)^\frac12 {\cal U}\,,
\end{align}
\end{subequations}
with  $O$  being an arbitrary  complex orthogonal matrix, ${\cal U}$  the \textit{unitary} matrix 
diagonalizing $\hat m_\nu$, $\hat m_{\rm heavy}$ 
the diagonal mass matrix of the heavy neutrinos and $\hat m_{\rm light}$ that of the light neutrinos. 
The entries of $\cal{U}$ are extracted from a global fit to atmospheric, reactor and solar neutrino oscillation 
data. Note that a possible non-unitarity of $\mathscr{U}$ is neglected in usual global fits, i.e. it is assumed that $\mathscr{U}={\cal U}$. The 
resulting  best fit values and $1\sigma$ ranges for the three mixing angles and the Dirac 
\textit{CP}-phase\footnote{The one sigma best fit region for the \textit{CP}-phase was determined in \cite{Fogli:2012ua} to be $0.77\,\pi\,-\, 1.36\,\pi$. However, the choice of the \textit{CP}-phase has no significant impact on our results.} read (see e.g.~\cite{Thomas})
\begin{subequations}
\begin{align}
\sin^{2}\theta_{12}&=0.30\pm 0.013\,,\\
\sin^{2}\theta_{23}&=0.41^{+0.037}_{-0.025} \,,\\
\sin^{2}\theta_{13}&=0.023\pm 0.0023\,,\\
\delta_{CP}&=300^{+66}_{-138}\,,
\end{align}
\end{subequations}
whereas the two Majorana phases are unknown and will be set to zero in the fit. The remaining degrees of freedom are the three masses of the heavy neutrinos and the entries of the matrix $O$. The latter can be parametrized by three complex angles. The freedom of choosing $O$ may be used to 
suppress the active-sterile mixing for one active neutrino flavour. Whether this suppression is possible or not depends on 
the structure of the light neutrino mass matrix, as can be inferred from \eqref{active-sterile-mixing}. 
\begin{table}[h]
\begin{tabular}{c|c|c|c}
\hline 
 & NH & IH & QD\\
\hline 
\hline 
$m_{1}\,{\rm (eV)}$ & $\sim 0$ & $4.85\cdot 10^{-2}$ & $\sim 0.1$\\
$m_{2}\,{\rm (eV)}$ & $8.660\cdot 10^{-3}$ & $4.93\cdot 10^{-2}$ & $\sim 0.1$\\
$m_{3}\,{\rm (eV)}$ & $4.97\cdot 10^{-2}$ & $\sim 0$ & $\sim 0.1$\\
\hline 
\end{tabular}
\caption{\label{NeutrinoMasses} Assumed masses of the light neutrinos for the normal 
(NH), inverted (IH) and quasi-degenerate (QD) neutrino mass spectra. }
\end{table}  
Since the absolute neutrino mass scale is unknown at present, we consider three 
possible neutrino mass spectra called NH, IH and QD which are defined in \tabl\ref{NeutrinoMasses}. In the case of the 
QD mass spectrum, the masses are constrained by the recent WMAP bound 
$\sum m_i\leq 0.44$ eV \cite{Hinshaw:2012fq}. For the numerical example  in the QD case 
we use $m_i\sim 0.1$ eV.

The minimization of the $\chi^2$ function is performed with a Monte Carlo method described in \app\ref{MetropolisAlgorithm}.
Let us note that this method can not find continuous solutions but falls randomly in the good fit points. 
For a given point in the parameter space we compute the $\mu\rightarrow e\gamma$ branching 
ratio and the effective electron neutrino mass $\langle m_{ee} \rangle$, relevant for the $0\nu \beta \beta$. For most 
of the points with a sizable active-sterile mixing one of these bounds is violated. To take 
the two bounds into account in the $\chi^2$ analysis we define 
\begin{subequations}
\label{chisqdef}
\begin{align}
\chi^2_{\mu\to e\gamma}&\equiv\left(\frac
{\text{BR}^{th}(\mu\to e\gamma)-\text{BR}^{exp}(\mu\to e\gamma)}
{\text{BR}^{exp}(\mu\to e\gamma)}\right)^2\nonumber\\
&\times\theta(\text{BR}^{th}(\mu\to e\gamma)-\text{BR}^{exp}(\mu\to e\gamma))\,,\\
\chi^2_{0\nu\beta\beta}&\equiv\left(\frac
{|\langle m^{th}_{ee}\rangle|-|\langle m^{exp}_{ee}\rangle |}
{|\langle m^{exp}_{ee}\rangle |}\right)^2\nonumber\\
&\times\theta(|\langle m^{th}_{ee}\rangle|-|\langle m^{exp}_{ee}\rangle |)\,,
\end{align}
\end{subequations}
where $\text{BR}^{th}(\m\rightarrow e\gamma)$ and  $\langle m^{th}_{ee} \rangle$ are the theoretical 
predictions computed at the chosen point of the parameter space, whereas 
$\text{BR}^{exp}(\m\rightarrow e\gamma)$ and  $|\langle m^{exp}_{ee} \rangle |$ are the 
corresponding $1\sigma$ experimental upper bounds. We use the theta step function to restrict the contributions to the total $\chi^2$ to cases when the theoretical prediction exceeds the one sigma exclusion limit. Additionally, we check whether the universality constraints, see \eqn\eqref{NonUniversConstr}, are fulfilled.
Finally, we compute the $S,T,U$ parameters and the corrected values of the electro-weak 
precision observables $O_i$, see \eqn\eqref{EWobservables}. The corresponding 
$\chi^2_{\text{EWPO}}$ value is calculated using 
\begin{align}
\label{chisqEWPOdef}
\chi^2_{\text{EWPO}}=\sum_i \frac{(O_i-O_{i,\text{SM}})^2}{(\delta O_i)^2+(\delta O_{i,\text{SM}})^2}\,,
\end{align}
where $O_{i,SM}$ denotes the predictions of the SM, $\delta O_{i,SM}$ are the theoretical
errors and $\delta O_i$ are the experimental errors, see \tabl\ref{ElectrowekObservables}.
Note that we neglect off-diagonal elements of the covariance matrix in these contributions.
The total $\chi^2$ is given by the sum of  \eqref{chisqdef} and \eqref{chisqEWPOdef}.

Using this definition we find that for points with negligibly small active-sterile mixing 
(`natural' see-saw) $\chi^2 \simeq \chi^2_{0} = 7.5$ ($7.6$ for the QD mass spectrum). This 
relatively large value of $\chi^2$ is induced primarily by the anomalies in the invisible 
$Z$-decay width, the NuTeV results and the deviation of the $W$ boson mass from the SM expectation. 
The slightly higher value for the QD mass spectrum stems from the fact that neutrino masses 
of the order of $0.1$ eV induce an effective electron neutrino mass $\langle m^{exp}_{ee} \rangle$ comparable to the current 
upper bound. 

To get a rough estimate of the goodness of the fit we compute the ratio of $\chi^2$ to the number 
of degrees of freedom. The $S$ and $U$ parameters are always very small and can be neglected 
in the fit. Therefore, the initial set of free parameters, the three masses and three complex 
angles, maps to four quantities: $\e_e$, $\e_\m$, $\e_\t$ and $T$. Since we fit six observables 
and two constraints, the  number of degrees of freedom is $(6+2)-4=4$ and 
$\chi^2_0/{\rm d.o.f} \approx 1.9$ for the case of negligible active-sterile mixing.  Below we perform a fit for each of the three light neutrino mass spectra to all EWPOs and constraints. In the following in all plots the full $\chi^2$ is presented. 

For all mass spectra we will show how the EWPOs in the best-fit regions are shifted with respect to the SM predictions. The question will be addressed whether a direct detection or an indirect signal of the new states is feasible in near future. To study this possibility the $\m \rightarrow e \gamma$ branching ratio, the $0\b\n\n$ decay rate and the strength of a collider signal will be estimated.     

\textit{Normal mass hierarchy.}
Let us first consider the normal mass hierarchy of the light neutrinos. The $\mu\rightarrow e\gamma$ 
branching ratio is suppressed in particular  if $\e_\m$  is small. As discussed above, if $S$ and $U$ are neglected 
then the $Z$ boson leptonic decay width  \eqref{Glept}  and the Weinberg  angle \eqref{sinTw} depend on the same combination of tree-level and one-loop 
corrections, $\e_e+\e_\m+2 \alpha_{em} T$. The reason is that in this approximation the shift of these 
observables is solely due to a shift of $G_\m$ with respect to $G_F$. 
\begin{figure}[h]
\includegraphics[width=\columnwidth]{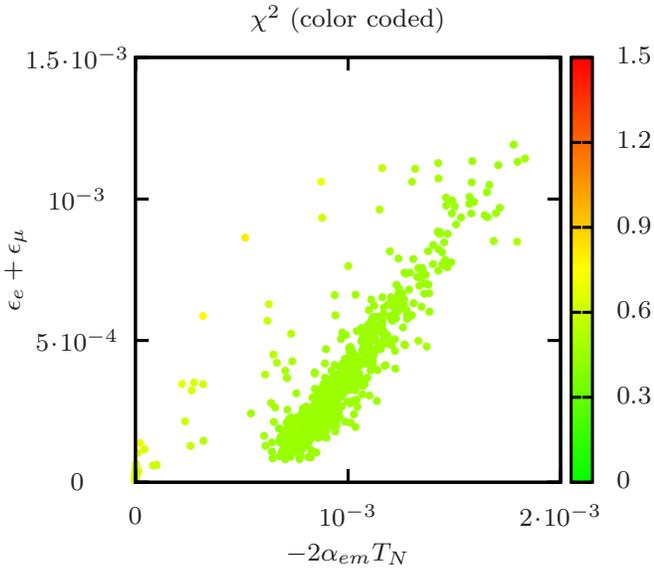}
\caption{\label{CancellationIllustrationTwoObs} $\chi^2$  as a function of $\e_e+\e_\m$ and $2\alpha_{em}T_N$ (NH), here only $\Gamma_\text{lept}$ and the Weinberg angle are fitted.}
\end{figure} 
In \fig\ref{CancellationIllustrationTwoObs} we show the fit to $\Gamma_\text{lept}$ and the Weinberg angle only. The plot shows the (approximately) linear relation between $\e_e+\e_\m$ and the $T$ parameter. The fact that the band does not start 
at the origin indicates that the two observables favour a slightly negative $T$ in general  even in a different new physics model. This plot demonstrates how a negative value of the T parameter can screen the contributions of the neutrino mixing  $\e_e+\e_\m$ to the above mentioned observables. We will see that this cancellation mechanism is still at work when more observables are considered.
\begin{figure}[h]
\includegraphics[width=\columnwidth]{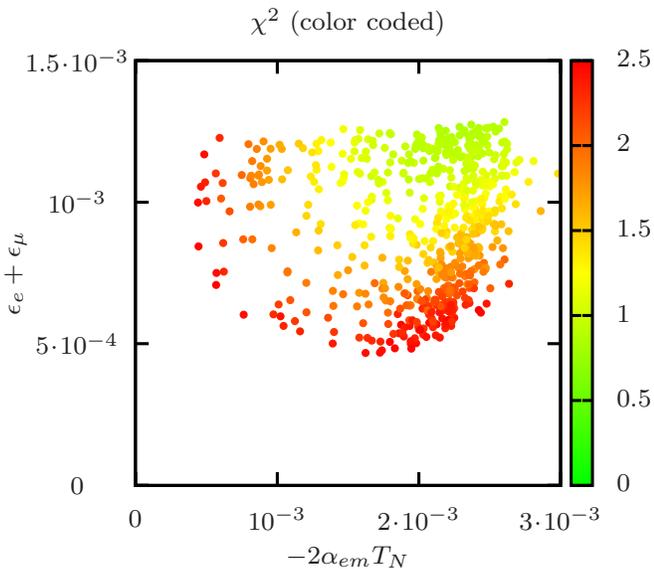}
\caption{\label{CancellationIllustration} $\chi^2$ for three d.o.f. as a function of $\e_e+\e_\m$ and $2\alpha_{em}T_N$ (NH). Here the $W$ boson mass is excluded from the fit.}
\end{figure}  
Large cancellations are present when the other EWPOs except the $W$ boson mass are taken into account. However, in this 
case the simple  linear dependence is violated due to the necessity of fitting the 
other observables, see  \fig\ref{CancellationIllustration}. The $W$ boson mass has been measured with a very high accuracy and a large negative $T$ would induce unacceptably large corrections to it. The fit excluding $M_W$ is of  interest  since an extended scalar sector appearing,
for instance, in the type-I+II see-saw extension of the SM will shift the $W$ mass and might absorb the additional negative $T$ contribution restricted in the type-I see-saw. 

Note that the following plots are two-dimensional projections of a multi-parameter scan and therefore points with rather 
different values of the color coded parameter can appear mixed on the plots.
\begin{figure}[h]
\includegraphics[width=\columnwidth]{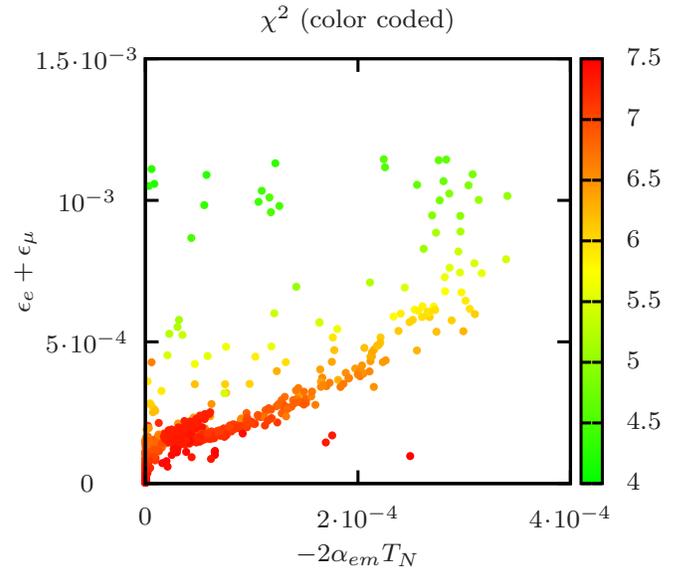}
\caption{\label{CancellationIllustrationWithmW} $\chi^2$ for four d.o.f. as a function of $\e_e+\e_\m$ and $2\alpha_{em}T_N$ (NH), fit to all EWPOs in \eqref{EWobservables}.}
\end{figure} 

In \fig\ref{CancellationIllustrationWithmW} we show how the cancellation pattern chan\-ges 
if all observables in \eqref{EWobservables} are included.  In this case for the best-fit points we have  $\chi^2\approx 4.0$  and 
a small negative $T$ is favored. The latter partially compensates the  small positive contribution
of $\e_e+\e_\m$ to $G_\mu$. 

\begin{figure}[h]
\includegraphics[width=\columnwidth]{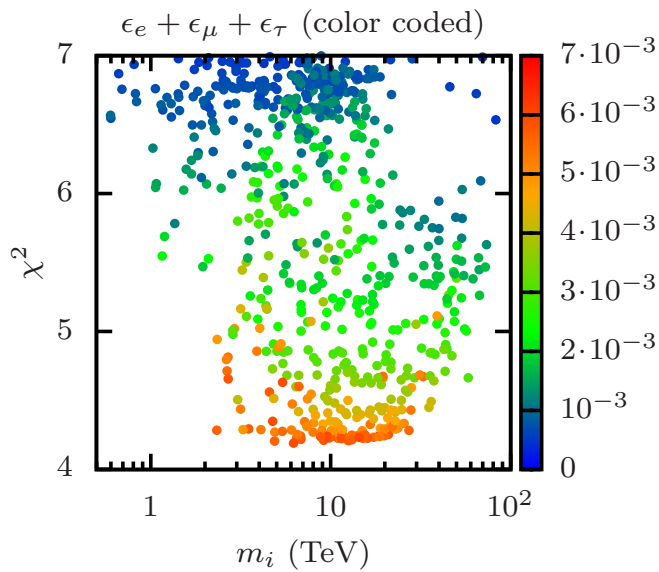}
\caption{\label{TeVSeeSawIllustrationB} $\e_e+\e_\m+\e_\t$ as a function of the lightest heavy neutrino mass
and $\chi^2$ for four d.o.f. (NH). Here $\e_\mu$ is suppressed.}
\end{figure}

As can be inferred from \fig\ref{TeVSeeSawIllustrationB}, at the best-fit points at least one the 
Majorana neutrinos is relatively light, around one TeV,
\begin{figure}[h]
\includegraphics[width=0.88\columnwidth]{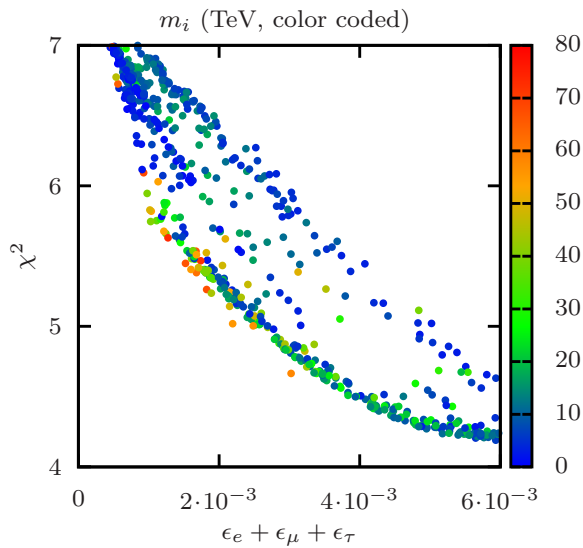}
\caption{\label{TeVSeeSawIllustrationA} 
The lightest heavy neutrino mass as a function of $\chi^2$ for four d.o.f. and $\e_e+\e_\m+\e_\t$ (NH). Here $\epsilon_\m$ is suppressed.}
\end{figure} 
and has a sizable coupling to at least one of the charged leptons, see \fig\ref{TeVSeeSawIllustrationA}. 
In other words, the current data favour the low-scale type-I see-saw with a considerable active-sterile 
mixing over the standard scenario (`natural' see-saw)  where this mixing is negligible and the masses of the 
heavy neutrinos are close to the GUT scale. On the other hand, as can be seen in a different projection of the 
same plot in \fig\ref{TeVSeeSawIllustrationB}, heavy neutrinos with masses below TeV scale are also disfavoured.
The reason is that for sizable  active-sterile mixing in this mass range experimental bounds are more stringent.  
As  already mentioned above, for a sizable active-sterile mixing the cancellation of the contributions to the light neutrino masses imposes constraints 
on the mass spectrum of the heavy ones. Due to requirements of suppressed $\e_{e/\mu}$ an active-sterile mixing pattern occurs where the first and third heavy mass eigenstates have comparable, sizable mixing to the active flavours while the mixing of the second is small.  We find  that the first and third heavy mass eigenstates have approximately equal masses whereas the mass of the remaining sterile neutrino is considerably larger. This mass pattern leads to a small negative $T$ parameter as discussed above.

At the best-fit points the deviation of the PMNS matrix from unitarity is of the order of
\begin{equation*}
\label{non-unitarity}
|(\mathscr{U U}^{\dagger})_{\alpha \beta}-\delta_{\alpha \beta}| \approx \left(
\begin{array}{ccc}
1 \cdot 10^{-3} & 1 \cdot 10^{-4}  & 2 \cdot 10^{-3}\\
1 \cdot 10^{-4} & 1 \cdot 10^{-5} & 2 \cdot 10^{-4}\\
2 \cdot 10^{-3} & 2 \cdot 10^{-4} & 5 \cdot 10^{-3}
\end{array}
\right)\,,
\end{equation*}
which is in excellent agreement with the bounds  derived in \cite{Antusch:2008tz}. This consistency check justifies our procedure and the applicability of the approximation in \eqref{active-sterile-mixing}.  
The corresponding (largest) zero-length transition probability is $P_{\nu_e\rightarrow \nu_\tau}\approx 7 \cdot 10^{-5}$. This value is 
too small to explain the current short-baseline anomalies \cite{ReactorAnomaly} but might contribute 
to the observed discrepancies.

\begin{figure}[h] 
\begin{center}
\includegraphics[width=\columnwidth]{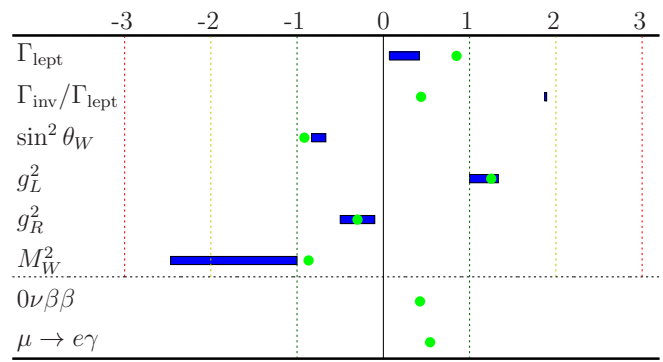}
\end{center}
\caption{\label{DevNormalSmallEmu} EWPOs calculated at the best-fit point  for NH and suppressed $\e_\mu$ (green dots) compared to the experimentally observed values, denoted by the zero line. The coloured lines stand for the respective experimental sigma deviations, thus the displacement of the predicted values form the observations is presented in units of the experimental error. Note that for the $0\nu\b\b$ and $\m \rightarrow e \gamma$ constraints we present only the one sigma exclusion limits.  The theoretical predictions of the SM with their theoretical uncertainties, see \tabl\ref{ElectrowekObservables}, are displayed as well (blue bars). (The best-fit point is at $M_1=20.3$ TeV, $M_2=14.1$ TeV, $M_3=21.0$ TeV, $\e_e=2.1\cdot 10^{-3}$, $\e_\m=3.0\cdot 10^{-6}$ and $\e_\t=4.5\cdot 10^{-3}$.)}
\end{figure}

In \fig\ref{DevNormalSmallEmu} we demonstrate how the EWPOs in the considered scenario are shifted from the values 
expected in the SM towards the experimentally observed ones. At the chosen best-fit point $\chi^2\approx 4.0$ and therefore the absolute 
improvement of the fit is $\Delta \chi^2\approx 3.5$ leading to a $\chi^2/{\rm d.o.f} \approx 1.0$. Some of the observables are shifted away from the experimentally measured value compared to the SM prediction. Nevertheless, they remain within the one sigma interval around the experimental results. In addition, observables for which the SM prediction is in tension with data are brought into the one sigma intervals around their experimental values. This leads to a global picture in which all observables agree on the one (or in the case of $g_L^2$ on the 1.2) sigma level with the experiments. This leads to a $\chi^2/{\rm d.o.f}$ of order one.  
The improvement stems primarily from $\e_\t$ which leaves $G_\mu$ unaffected (and thus equal to $G_F$) but suppresses the invisible 
$Z$-decay width. Furthermore, $\e_e$ shifts the $W$ boson mass towards the measured value. The NuTeV observables 
are not significantly changed.  The last two rows in \fig\ref{DevNormalSmallEmu} present the ratios of the induced 
effective electron neutrino mass $\langle m_{ee} \rangle$ and the $\mu\rightarrow e\gamma$ branching ratio to the corresponding experimental 
bounds. 
\begin{figure}[h]
\includegraphics[width=\columnwidth]{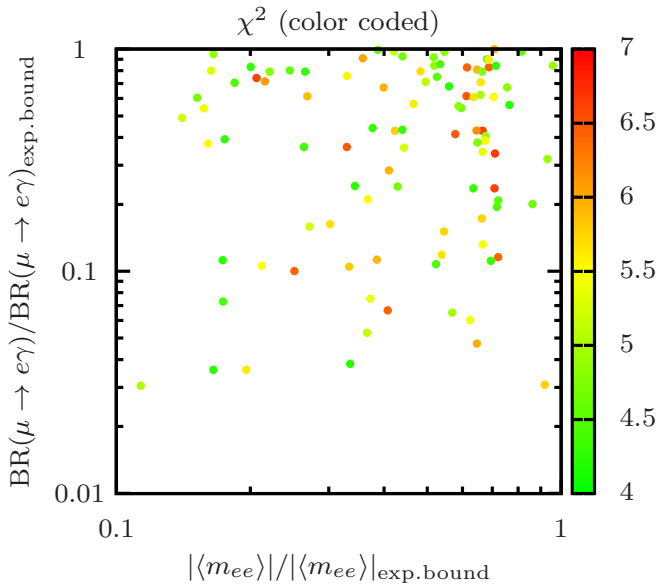}
\caption{\label{NH_Smu_RP} $\chi^2$  for four d.o.f. as a function of the ratios of the  $\mu\rightarrow e\gamma$ branching 
ratio and $|\langle m_{ee} \rangle |$ to the corresponding experimental bounds (NH). Here $\e_\mu$ is suppressed.}
\end{figure} 
It is interesting to note that for the normal mass hierarchy there are many points in the parameters space 
with values of these two quantities close to the threshold of the current experimental sensitivity, see 
\fig\ref{NH_Smu_RP}. Here the predictions for the $\mu\rightarrow e\gamma$ branching ratio and $|\langle m_{ee} \rangle |$ of the points in the parameter space with significant improvement of the $\chi^2$ are compared to current experimental limits.  
Knowing the masses of the heavy neutrinos and their couplings  to the charged leptons we can also 
estimate the cross-section of the direct detection process depicted in \fig\ref{DirectNProcess}. 
For all best-fit points the resulting value of $|\sum_i \mathbf{U}^2_{\alpha i}\,m^{-1}_i|$
turns out to be at most of the order of $10^{-5}$ and thus beyond the LHC reach.

\begin{figure}[h]
\begin{center}
\includegraphics[width=\columnwidth]{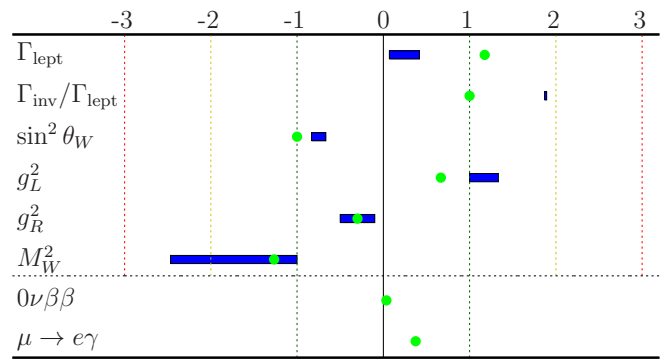}
\end{center}
\caption{\label{DevNormalSmallEe} This is the analogue of \fig\ref{DevNormalSmallEmu} but for the case 
of suppressed $\e_e$ in NH. At the best-fit point $M_1=4.1$ TeV, $M_2=161.0$ TeV, 
$M_3=7.1$ TeV, $\e_e=1.0\cdot 10^{-6}$, $\e_\m=1.5\cdot 10^{-3}$ and $\e_\t=1.2\cdot 10^{-3}$.}
\end{figure}

The $\mu\rightarrow e\gamma$ branching ratio is also suppressed if $\e_e$  
is small. In this case we obtain a slightly worse best-fit value of $\chi^2 \approx 4.8$ which 
corresponds to $\chi^2/{\rm d.o.f} \approx 1.2$. As compared 
to the previous case, in the best-fit points the invisible $Z$-decay width is less suppressed and 
the shift in the $W$ boson mass is smaller, see \fig\ref{DevNormalSmallEe}. On the other hand, $g_L^2$ is 
shifted towards the value measured by the NuTeV collaboration and thus brought in agreement with the data within the one sigma interval.

\begin{figure}[h]
\begin{center}
\includegraphics[width=\columnwidth]{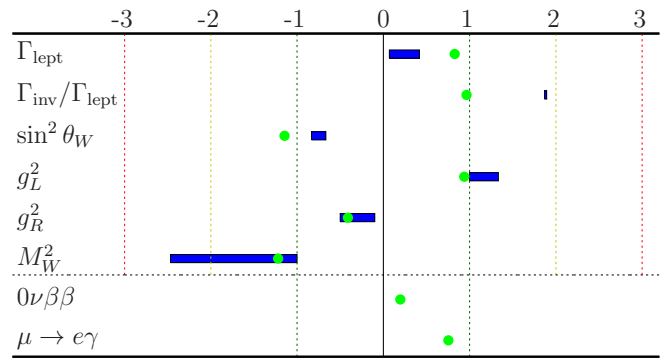}
\end{center}
\caption{\label{DevInverseSmallEe}This is the analogue of \fig\ref{DevNormalSmallEmu} but for the case of IH and suppressed $\e_e$. At the best-fit point $M_1=551$ GeV, $M_2=242$ GeV, $M_3=469$ GeV, $\e_e=3\cdot 10^{-6}$, $\e_\m=1.58\cdot 10^{-3}$ and $\e_\t=1.1\cdot 10^{-3}$.}
\end{figure}

\begin{figure}[h]
\includegraphics[width=\columnwidth]{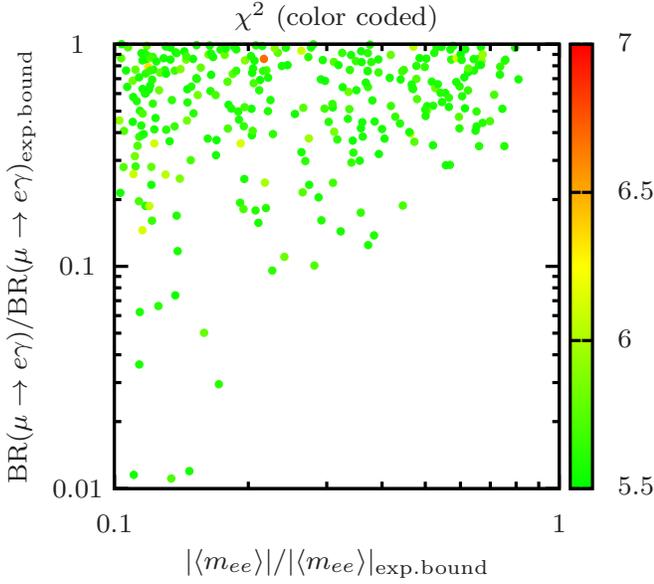}
\caption{\label{IH_RP} $\chi^2$ for four d.o.f. as a function of the ratios of the  $\mu\rightarrow e\gamma$ branching 
ratio and $|\langle m_{ee} \rangle |$ to the corresponding experimental bounds in the case of IH and suppressed $\e_e$.}
\end{figure}

\begin{figure}[h]
\includegraphics[width=\columnwidth]{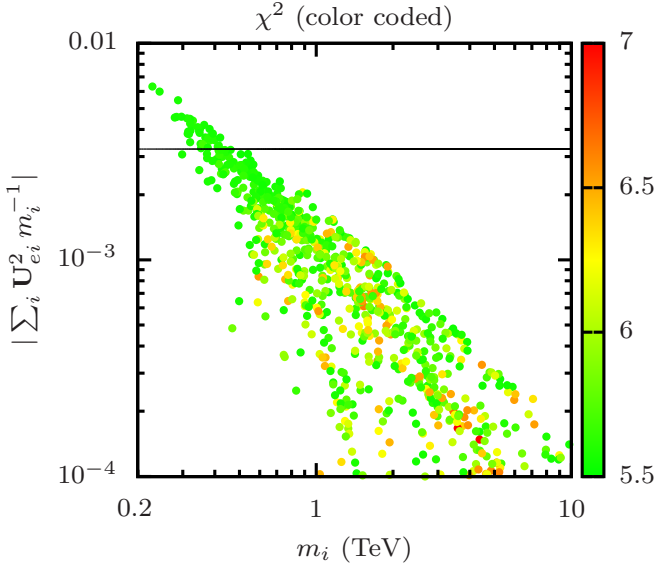}
\caption{\label{IH_LHC} $\chi^2$ for four d.o.f. as a function of mass of the heavy neutrino giving the leading contribution 
and co\-u\-pling to the muon in the case of IH and suppressed $\e_e$. The region above the solid line can be tested by the LHC after the upgrade 
to 14 TeV if the anticipated luminosity $\sim 400 \text{fb}^{-1}$ is reached.}
\end{figure}

\textit{Inverted mass hierarchy.}
For the inverted mass hierarchy scenario only the option of negligible $\e_e$ 
can be realized and results in best fit values of $\chi^2 \approx 5.5$ and $\chi^2/{\rm d.o.f} \approx 1.4$. 
The EWPOs of the best-fit points are influenced such that the predicted invisible $Z$-decay width is 
brought within the one sigma interval to the experimental value, see \fig\ref{DevInverseSmallEe}. 
The $W$ boson mass is shifted in the right direction but less significantly than in the normal hierarchy 
scenario. On the other hand, $g_L^2$ is shifted towards the value measured by the NuTeV collaboration 
and brought in good agreement with the data. 
As can be inferred from \fig\ref{IH_RP}, both the effective electron neutrino mass $\langle m_{ee} \rangle$ and the $\mu\rightarrow e\gamma$
branching ratio are in large parts of the parameter space close to the current experimental sensitivity. 
Furthermore, for a small fraction of the best-fit points, see \fig\ref{IH_LHC},
there is a chance of observing the heavy Majorana neutrinos at the LHC after the planned upgrade to 
$\sqrt{s}=14$ TeV.

\begin{figure}[h]
\begin{center}
\includegraphics[width=\columnwidth]{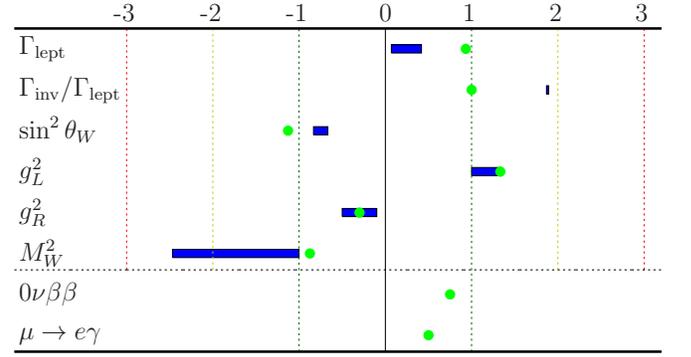}
\end{center}
\caption{\label{DevQDSmallEmu}This is the analogue of \fig\ref{DevNormalSmallEmu} but for the case of 
the QD mass spectrum and suppressed $\e_\mu$. At the best-fit point $M_1=14.5$ TeV, 
$M_2=10.5$ TeV, $M_3=14.9$ TeV, $\e_e=1.5\cdot 10^{-3}$, $\e_\m=2.9 \cdot 10^{-6}$ and $\e_\t=2.4\cdot 10^{-3}$.}
\end{figure}

\begin{figure}[h]
\includegraphics[width=\columnwidth]{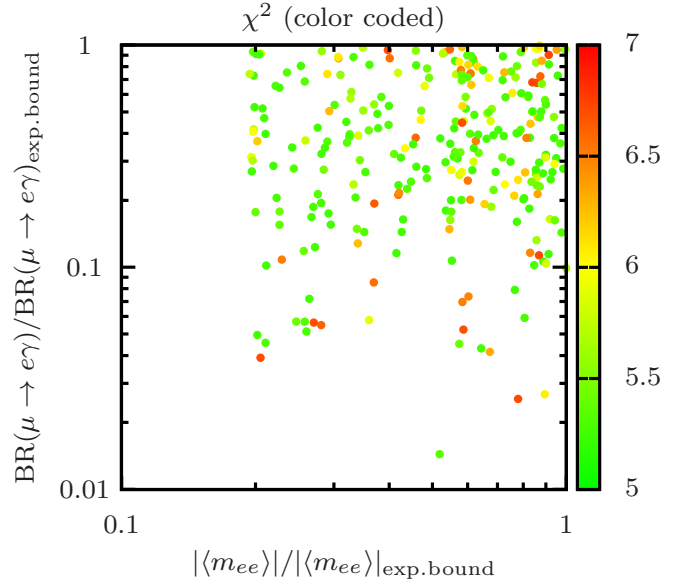}
\caption{\label{QD_RP} $\chi^2$ for four d.o.f. as a function of the ratios of the  $\mu\rightarrow e\gamma$ branching 
ratio and $|\langle m_{ee} \rangle |$ to the corresponding experimental bounds in case of quasi-degenerate mass spectrum and suppressed $\e_\mu$.}
\end{figure}

\textit{Quasi-degenerate mass spectrum.}
For the quasi-degenerate mass spectrum it is quite difficult to find points with sizable $\epsilon_\mu$ 
that pass the $\mu\rightarrow e\gamma$ constraint. On the other hand, there are points with small 
$\epsilon_\m$ and sizable $\epsilon_e$ and $\epsilon_\t$ which satisfy the $\mu\rightarrow e\gamma$, 
neutrinoless double-beta decay  and universality constraints. For the best-fit points $\chi^2 \approx 5$ 
leading to an improvement of $\Delta \chi^2\approx 2.6$ and to $\chi^2/{\rm d.o.f} \approx 1.25$.  
At the best-fit points the shift of  the invisible $Z$-decay width is smaller than in the normal 
hierarchy scenario, but still the prediction moves into the one sigma interval of the observations. The shift in the $W$ boson brings prediction and experiment in excellent agreement, 
see \fig\ref{DevQDSmallEmu}. The neutrino scattering observable $g_L^2$ is not significantly influenced in 
comparison to the standard model expectation. 
As can be inferred from \fig\ref{QD_RP} both the effective mass of the electron neutrino and $\mu\rightarrow e\gamma$
branching ratio here can be close to the exclusion limit. 
Note that the active neutrinos induce a sizable contribution to the effective electron neutrino mass $|\langle m_{ee} \rangle|$ in this 
case  and can constructively  or destructively interfere with the heavy neutrino contributions depending on 
the Majorana phases of the active neutrinos. 

For all the best-fit points the resulting value of $|\sum_i \mathbf{U}^2_{\alpha i}\,m^{-1}_i|$
turns out to be at most of the order of $10^{-5}$ and thus beyond the LHC reach.

It has been found that for all three mass spectra the current data favour a type-I see-saw with considerable active-sterile mixing over the a scenario where the mixing is negligible. For all three cases there are many points in the parameter space which generate signals in rare decay experiments close to the current sensitivities. For the IH a part of the parameter space can be tested at the LHC after the next upgrade.

\section{\label{Summary} Summary and outlook}

In this work we have studied the impact of  TeV-scale sterile neutrinos with a sizable active-sterile 
mixing on electro-weak precision observables and lepton number and flavour violating decays in the framework of the type-I see-saw extension of the SM. 
The  active-sterile mixing can in particular improve the global fit and reduce some anomalies in the experimental data: first, it results 
in a decreased value of the invisible $Z$ boson decay width, which is preferred 
by the current data; second, it slightly increases  the mass of the $W$ boson and brings 
it in the one sigma interval around the observed value. Furthermore, in the case when  $\e_\mu$ is not suppressed, the model affects the charged-to-neutral 
current ratio for neutrino scattering and brings $g_{L}$ in agreement with the values observed by the NuTeV experiment.

In our analysis we have taken into account tree-level contributions of sterile neutrinos as well as their one-loop effects on gauge boson propagators. 
The tree-level contributions enter via non-unitarity of the PMNS mixing matrix, whereas 
the loop-contributions modify the propagators of the gauge bosons and can be taken into account in form of the $S,T,U$ parameters. 
The oblique corrections can play an important role. In particular we have demonstrated that a sizable 
negative $T$ parameter `screens'  the effect of tree-level non-unitarity on the 
Fermi constant. This cancellation mechanism can reconcile large active-sterile 
mixing with the current observations. 
As compared to previous analyses we have derived explicit 
formulae for the oblique corrections in terms of the Majorana masses and mixings and therefore studied 
a full and self-consistent model. 
We have performed a numerical fit to the electro-weak precision observables taking 
into account constraints from the non-observation of $\mu\rightarrow e\gamma$ and neutrinoless 
double-beta decay processes as well as constraints on lepton non-universality. Since the active neutrino mass hierarchy is unknown at present, 
we have considered  the cases of normal, inverted and quasi-degenerate mass spectra. In all three cases 
regions of the parameter space with a sizable active-sterile mixing provide a better overall fit 
to the data than regions where it is negligible. In other words, the current data favour the 
low-energy type-I see-saw with a considerable active-sterile mixing over the standard scenario 
(`natural' see-saw)  where this mixing is negligible. 
Together with the derivation of expressions for the oblique corrections in the type-I see-saw this finding is one of the main results of the present work. 

The $\chi^2$ is lowest for the normal hierarchy scenario and slightly higher for the inverted hierarchy and the case of the quasi-degenerate mass spectrum. We have also studied the experimental signatures of the best-fit points for each mass hierarchy. 
For all mass spectra the effective electron neutrino 
mass $\langle m_{ee} \rangle$ -- the quantity relevant for the neutrinoless double-beta decay --  reaches at  many best-fit points 
values that are close to the sensitivity threshold of current experiments.   The expected active-sterile mixing is of the order of $\mathbf{U}_{(e/\mu)i}^2 \approx 10^{-3}$ in all three scenarios. The lightest sterile neutrino mass for normal- and quasi-degenerate mass spectra is around 2 TeV. This ranges might become accessible at electron-proton colliders with energies above 6 TeV. For the inverted 
hierarchy the Majorana masses can be as low as 300 GeV. Thus, given the predicted mixing, 
this scenario can be partially tested at the LHC after the 14 TeV upgrade.

The fit might improve in a type I+II see-saw extension of the SM, since the VEV of the additional Higgs triplet also contributes to the $W$ boson mass. Thus the negative $T$ parameter can be larger and the cancellation mechanism more efficient.  This extension will be studied in a forthcoming paper.

\section*{Acknowledgements}
The work of A.K. has been supported by the German Science Foundation (DFG) under Grant 
KA-3274/1-1 ``Systematic analysis of baryogenesis in non-equilibrium quantum field theory''. 
J.S. acknowledges support by the IMPRS for Precision Tests of Fundamental Symmetries.
The authors would like to thank M. Blennow, T. Schwetz , J. Heeck, J. Barry, W. Rodejohann and S. Antusch 
for helpful discussions. 

\appendix
\section{\label{DefOfSTU}Oblique corrections}

The formalism of oblique corrections was developed to study   new physics which 
affects the observables only via corrections to the propagators of the gauge bosons
\cite{STUOriginal,BurgessSTU}. The Lorentz structure of the corresponding self-ener\-gies is
\begin{align}
\Pi_{ab}^{\m\n}(q^{2}) & =\Pi_{ab}(q^{2})g^{\m\n}+(q^{\m}q^{\n}\,\,\text{terms})\,,
\end{align}
where $a,b=\gamma,W^{\pm},Z$. Since in the considered processes all external fermions 
are light, the contributions of the $q^{\m}q^{\n}$ terms is proportional to $m_f/M_{W,Z}$
and can safely be neglected. For the same reason, contributions of the box 
diagrams as well as the vertex corrections stemming from the new physics are negligible.
Therefore, it is sufficient to consider only $\Pi_{ab}$. The latter can be represented 
in the form 
\begin{equation}
\Pi_{ab}(q^{2})=\Pi_{ab}^{SM}(q^{2})+\v\Pi_{ab}(q^{2})\,,
\end{equation}
where $(ab)=(\g\g),\,(Z\g),\,(ZZ), (WW)$. It follows from the Ward identity that  
$\v\Pi_{\g\g}(0)=\v\Pi_{\g Z}(0)=0$  which further constrains the number of 
independent quantities. It can be shown  that for heavy new physics there are 
only three independent combinations of the above quantities that enter the 
expressions for the electro-weak observables \cite{STUOriginal}. These are 
denoted by $S$, $T$ and $U$ and read \cite{STUOriginal,Kniehl:1992ez}:
\begin{subequations}
\label{STUdefinition}
\begin{align}
\label{eq:S}
S&=\frac{4s_{w}^{2}c_{w}^{2}}{M_{Z}^{2}}\Bigl[\hat{\Pi}_{ZZ}(0)
+\hat{\Pi}_{\g\g}(M_{Z}^{2})\nonumber\\
&\hspace{25mm}-\frac{c_{w}^{2}-s_{w}^{2}}{c_{w}s_{w}}\hat{\Pi}_{Z\g}(M_{Z}^{2})\Bigr]\,,\\
\label{eq:T}
T&=\frac{\hat{\Pi}_{ZZ}(0)}{M_{Z}^{2}}-\frac{\hat{\Pi}_{WW}(0)}{M_{W}^{2}}\,,\\
\label{eq:U}
U&=4s_{w}^{2}c_{w}^{2}\biggl[\frac{1}{c_{w}^{2}}\frac{\hat{\Pi}_{WW}(0)}{M_{W}^{2}}-
\frac{\hat{\Pi}_{ZZ}(0)}{M_{Z}^{2}}+\nonumber\\
&\hspace{5mm}+\frac{s_{w}^{2}}{c_{w}^{2}}\frac{\hat{\Pi}_{\g\g}(M_{Z}^{2})}{M_{Z}^{2}}-
\frac{s_{w}}{c_{w}}\frac{2\hat{\Pi}_{Z\g}(M_{Z}^{2})}{M_{Z}^{2}}\biggr]\,.
\end{align}
\end{subequations}
The hats denote self-energies renormalized using the on-shell renormalization 
condition:
\begin{align}
\label{OSconditions}
\Re\, \hat{\Pi}_{WW}(M_{W}^{2})&=\Re\, \hat{\Pi}_{ZZ}(M_{Z}^{2})\nonumber\\
&=\hat{\Pi}_{Z\g}(0)=\hat{\Pi}_{\g\g}(0)=0\,.
\end{align}
Explicit formulae for the renormalized self-energies in terms of the 
unrenormalized ones are \cite{Kniehl:1992ez}:
\begin{subequations}
\label{OnShellSE}
\begin{align}
\hat{\Pi}_{WW}(q^{2})&=\Pi_{WW}(q^{2})-\Pi_{WW}(M_{W}^{2})\nonumber\\
+&\left(q^{2}-M_{W}^{2}\right)\left[\left(\frac{c^{2}}{s^{2}}\right)R-\Pi'_{\g\g}(0)\right]\,,\\
\hat{\Pi}_{ZZ}(q^{2})&=\Pi_{ZZ}(q^{2})-\Pi_{ZZ}(M_{w}^{2})\nonumber\\
+&\left(q^{2}-M_{Z}^{2}\right)\left[\left(\frac{c^{2}}{s^{2}}-1\right)R-\Pi'_{\g\g}(0)\right]\,,\\
\hat{\Pi}_{Z\g}(q^{2})&=\Pi_{Z\g}(q^{2})-\Pi_{Z\g}(0)-q^{2}\left(\frac{c^{2}}{s^{2}}\right)R\,,\\
\hat{\Pi}_{\g\g}(q^{2})&=\Pi_{\g\g}(q^{2})-q^{2}\Pi'_{\g\g}(0)\,,
\end{align}
\end{subequations}
where
\begin{align}
R=\frac{\Pi_{ZZ}(M_{Z}^{2})}{M_{Z}^{2}}-\frac{\Pi_{WW}(M_{W}^{2})}{M_{W}^{2}}-
2\frac{s_{w}}{c_{w}}\frac{\Pi_{Z\g}(0)}{M_{Z}^{2}}\,.
\end{align}
Substituting \eqref{OnShellSE} into \eqref{STUdefinition} and taking into account 
that $\Pi_{Z\g}(0)=0$ we find: 
\begin{subequations}
\label{STUunrenormalizedSE}
\begin{align}
\label{SunrenormalizedSE}
S & =\frac{4s_{w}^{2}c_{w}^{2}}{M_{Z}^{2}}\biggl[
\Pi_{ZZ}(0)-\Pi_{ZZ}(M_{Z}^{2})+\Pi_{\g\g}(M_{Z}^{2}) \nonumber \\
& \hspace{35mm} -\frac{c_{w}^{2}-s_{w}^{2}}{c_{w}s_{w}}\Pi_{Z\g}(M_{Z}^{2})\biggr]\,, \\
\label{TunrenormalizedSE}
T & =\frac{\Pi_{ZZ}(0)}{M_{Z}^{2}}-\frac{\Pi_{WW}(0)}{M_{W}^{2}}\,, \\
\label{UunrenormalizedSE}
U & =4s_{w}^{2}c_{w}^{2}\biggl[\frac{\Pi_{WW}(0)-\Pi_{WW}(M_{W}^{2})}{c_{w}^{2} M_{W}^{2}}  \nonumber \\
&- \frac{\Pi_{ZZ}(0)-\Pi_{ZZ}(M_{Z}^{2})}{M_{Z}^{2}} +\frac{s_{w}^{2}}{c_{w}^{2}}\frac{\Pi_{\g\g}(M_{Z}^{2})}{M_{Z}^{2}}\nonumber\\
&\hspace{40mm}-2\frac{s_{w}}{c_{w}}\frac{\Pi_{Z\g}(M_{Z}^{2})}{M_{Z}^{2}}\biggr]\,.
\end{align}
\end{subequations}
Note that only for the $T$ parameter the expression in terms of the unrenormalized 
self-energies has the same form as in terms of the renormalized ones, compare \eqref{eq:T} 
and \eqref{TunrenormalizedSE}.

\section{\label{STUcalculation}Calculation of \textit{S,T,U}}

\begin{figure}[h]
\begin{center}
\includegraphics[width=0.5\columnwidth]{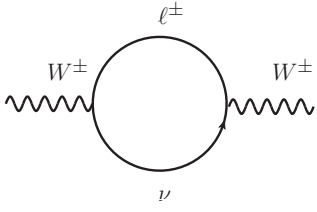}
\end{center}
\caption{\label{PiWW}Contribution of the charged leptons and neutrinos to $\Pi_{WW}$.}
\end{figure}

\begin{figure}[h]
\begin{center}
\includegraphics[width=\columnwidth]{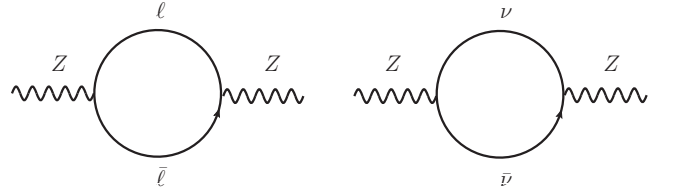}
\end{center}
\caption{\label{PiZZ}Contribution of the charged leptons and neutrinos to $\Pi_{ZZ}$.}
\end{figure}

To evaluate the contribution of $n$ Majorana neutrinos to the $S,T,U$ parameters defined in 
\app\ref{DefOfSTU} we need to calculate the self-energies entering \eqref{STUunrenormalizedSE}. 
\begin{figure}[h]
\begin{center}
\includegraphics[width=\columnwidth]{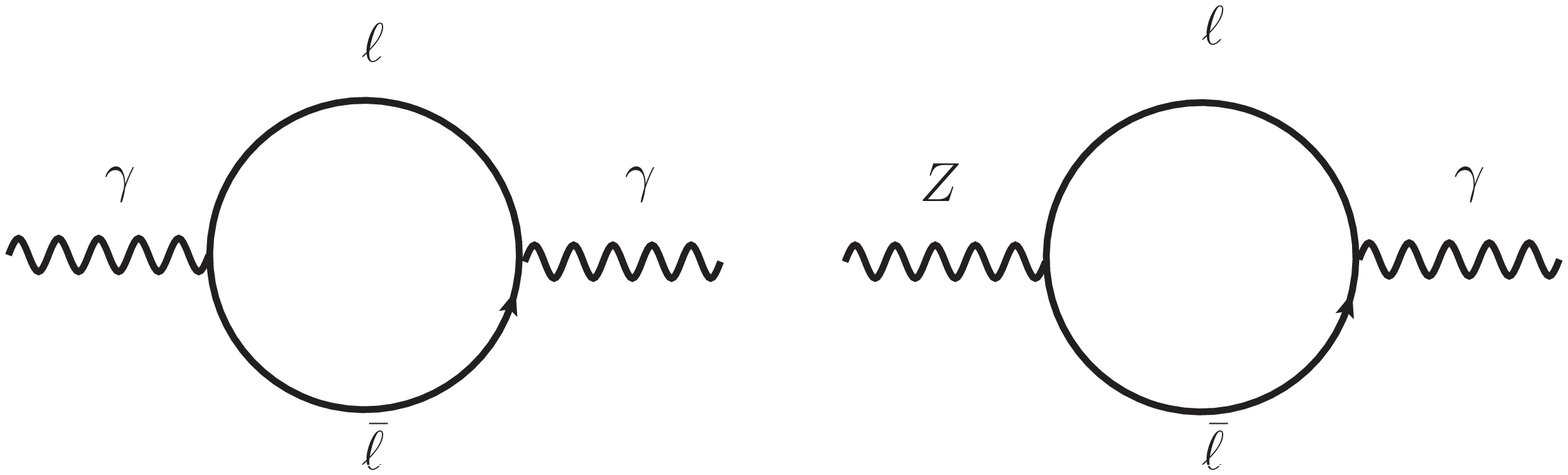}
\end{center}
\caption{\label{PiZgPigg}Contribution of the charged leptons to $\Pi_{\gamma\gamma}$ and $\Pi_{Z\gamma}$ at one-loop 
level.}
\end{figure} 
Only charged leptons contribute to $\Pi_{Z\g}$ and $\Pi_{\g\g}$, see \fig\ref{PiZgPigg}.  
The resulting  self-energies are the same as in the SM:
\begin{subequations}
\label{PiZggg}
\begin{align}
\Pi^{\mu\nu}_{\g\g}(q^2)&=\left(g^{\mu\nu}-\frac{q^\mu q^\nu}{q^2}\right)\frac{-e^{2}}{4\pi^{2}}\sum_{\a}\bigl[ Q(q^{2},m^2_{\a},m^2_{\a})\nonumber\\
&\hspace{20mm}-m_{\a}^{2}B_0(q^{2},m^2_{\a},m^2_{\a})\bigr]\,,\\
\Pi^{\mu\nu}_{Z\g}(q^2)&=\frac{\left(4s_{w}^{2}-1\right)}{4c_{w}s_{w}}\Pi^{\mu\nu}_{\g\g}(q^2)\,,
\end{align}
\end{subequations}
where $m_\alpha$ denote the masses of the charged leptons. To shor\-ten the notation in \eqref{PiZggg}
we have introduced 
\begin{align}
\label{Qdef}
Q(q^2,&m^2_1,m^2_2)\equiv (D-2)B_{22}(q^2,m_1^2,m_2^2)\nonumber\\
&+q^2\bigl[ 
B_1(q^2,m_1^2,m_2^2)+B_{21}(q^2,m_1^2,m_2^2)
\bigr]\,,
\end{align}
where $B_0$, $B_1$, $B_{21}$ and $B_{22}$ are the usual one-loop functions
\cite{Passarino:1978jh}, $D\equiv 4-2\epsilon$  and $\epsilon\rightarrow 0$.
The one-loop contribution of the charged leptons and neutrinos to $\Pi_{WW}$ is 
presented in \fig\ref{PiWW}.
Note that due to the nonzero active-sterile mixing the heavy neutrinos can 
also run in the loop. In position space the resulting self-energy is given by:
\begin{align}
\label{PiWWcoord}
\Pi^{\m\n}_{WW}(x,y)&=-\frac{e^{2}}{2s_{w}^{2}}\sum_{\a,i} |\mathbf{U}_{\a i}|^{2} \nonumber \\
&\times\bigl\langle T\bigl[ \bar{\n}_{i}(x)\g^{\n}P_{L}\ell^{\a}(x)\bar{\ell}_{\a}(y)\g^{\m}P_{L}\n^{i}(y)+ \nonumber \\
& +\bar{\ell}_{\a}(x)\g^{\m}P_{L}\n^{i}(x)\bar{\n}_{i}(y)\g^{\n}P_{L}\ell^{\a}(y)\bigr] \bigr\rangle\,.
\end{align}
Since we deal with Majorana fermions, to evaluate \eqref{PiWWcoord} it is convenient 
to use the two component notation for the spinors, see \app\ref{TwoCompNotation} for more details. 
The self-energy then takes the form:
\begin{align}
\label{PiWWcoordtwocomp}
\Pi^{\m\n}_{WW}(x,y)&=-\frac{e^{2}}{2s_{w}^{2}}\sum_{\a,i}|\mathbf{U}_{\a i}|^{2} \nonumber \\
& \times \bigl\langle T\bigl[\chi_{i}(x)\bar{\sigma}^{\m}\ell_{L}^{\a}(x)
\bar{\ell}_{L}^{\a}(y)\bar{\sigma}^{\n}\chi^{i}(y) \nonumber \\
&+\bar{\ell}_{L}^{\a}(x)\bar{\sigma}^{\m}\chi^{i}(x)\chi_{i}(y)
\bar{\sigma}^{\n}\ell_{L}^{\a}(y)\bigr] \bigr\rangle\,,
\end{align}
where $\ell_{L}$ denotes left-handed component of the charged-lepton field in the 
two-component notation. To evaluate \eqref{PiWWcoordtwocomp}  we need to consider 
all possible contractions of the field operators. Using the Fourier-representation of 
the propagators we obtain:
\begin{align}
\label{PiWWinterm}
\Pi^{\m\n}_{WW}(q^2)&=-i\frac{e^{2}}{2s_{w}^{2}}\sum_{\a,i}|\mathbf{U}_{\a i}|^{2}\nonumber\\
&\times \int\frac{d^{4}p}{(2\pi)^{4}} \frac{d^{4}k}{(2\pi)^{4}} (2\pi)^{4} \delta(q-k+p)\nonumber\\
&\times \frac{\tr\left[\slashed{k}\g^{\m}\slashed{p}\g^{\n}\right]}{\left(p^{2}-m_{i}^{2}+
i\varepsilon\right)\left(k^{2}-m_{\alpha}^{2}+i\varepsilon\right)}\,.
\end{align}
Note that since only the left-handed component of the charged field runs in the loop, 
only `kinetic' contraction is possible. This is reflected by the fact that the numerator 
of \eqref{PiWWinterm} contains only the momenta $\slashed{k}$ and $\slashed{p}$ of the 
intermediate states. Taking the trace and using the definitions of the one-loop functions 
we finally arrive at:
\begin{align}
\label{PiWWappendix}
\Pi^{\m\n}_{WW}(q^2)&=-\frac{e^{2}}{16\pi^{2}s_{w}^{2}}\sum_{\a,i}|\mathbf{U}_{\a i}|^{2}
\bigl[g_{\m\n}Q(q^{2},m_{i}^{2},m_{\a}^{2})\nonumber\\
&\hspace{20mm}-q_{\m}q_{\n}\,P(q^{2},m_{i}^{2},m_{\a}^{2})\bigr]\,,
\end{align}
where $Q$ has been defined above \eqref{Qdef} and 
\begin{align*}
P(q^{2},m_1^2,m_2^2)\equiv 2B_{21}(q^{2},m_1^2,m_2^2)+2B_{1}(q^{2},m_1^2,m_2^2)\,.
\end{align*}
For the $Z$ boson self-energy we need to consider two diagrams with the neutral or charged 
states propagating in the loop, see \fig\ref{PiZZ}.
In this case both `kinetic'  and `mass' contractions are possible. In complete analogy 
to \eqref{PiWWappendix} the kinetic contraction of the two neutrino lines results in:
\begin{align}
\label{PiZZ1}
\Pi^{\m\n}_{ZZ(1)}&(q^2)=-\frac{e^{2}}{32\pi^{2}s_w^2 c_w^2}
\sum_{ij}\sum_{\a\b}\mathbf{U}_{i\a}^{\dagger}\mathbf{U}_{\a j}\mathbf{U}_{j\b}^{\dagger}\mathbf{U}_{\b i}\nonumber\\
\times &\bigl[\,g_{\m\n}Q(q^{2},m_{i}^{2},m_{j}^{2})-q_{\m}q_{\n}\,P(q^{2},m_{i}^{2},m_{j}^{2})\bigr]\,.
\end{align}
Since the intermediate neutrinos are Majorana particles the `mass' contraction is 
also possible:
\begin{align}
\label{PiZZ2}
\Pi^{\m\n}_{ZZ(2)}(q^2)&=-\frac{e^{2}}{32\pi^{2}s_w^2 c_w^2}
\sum_{ij}\sum_{\a\b}\mathbf{U}_{i\a}^{\dagger}\mathbf{U}_{\a j}\mathbf{U}_{i\b}^{\dagger}\mathbf{U}_{\b j}\nonumber\\
&\times  \bigl[\,g_{\m\n} \, m_i m_j B_0(q^{2},m_{i}^{2},m_{j}^{2})\bigr]\,.
\end{align}
This contribution vanishes for vanishing Majorana mass as it should. Note also that
the flavor structures of \eqref{PiZZ2} are slightly different compared to \eqref{PiZZ1}.

Since the left- and right-handed charged leptons couple to $Z$ with the strengths $1-2s_w^2$ and $2s_w^2$
respectively, the contribution of the kinetic contraction reads: 
\begin{align}
\label{PiZZ3}
\Pi^{\m\n}_{ZZ(3)}&(q^2)=-\frac{e^{2}}{32\pi^{2}s_w^2 c_w^2}\sum_\a
\bigl[(1-2s_w^2)^2+(2s_w^2)^2\bigr]\nonumber\\
\times \bigl[&\,g_{\m\n}Q(q^{2},m_{\a}^{2},m_{\a}^{2})-q_{\m}q_{\n}\,P(q^{2},m_{\a}^{2},m_{\a}^{2})\bigr]\,.
\end{align}
Since charged leptons are Dirac particles  the mass term appears only for contractions 
of the left- and right-handed components and is therefore proportional to a product of 
the two couplings:
\begin{align}
\label{PiZZ4}
\Pi^{\m\n}_{ZZ(4)}(q^2)&=-\frac{e^{2}}{32\pi^{2}s_w^2 c_w^2}
\sum_{\a} 4s_w^2(1-2s_w^2) \nonumber\\
&\times  \bigl[\,g_{\m\n} \, m^2_\a B_0(q^{2},m_{\a}^{2},m_{\a}^{2})\bigr]\,.
\end{align}
The total contribution to the $Z$ boson self-energy is given by the sum of \eqns\eqref{PiZZ1}-\eqref{PiZZ4}.

Substituting the terms proportional to $g^{\m\n}$  of $\Pi^{\m\n}_{WW}$ and $\Pi^{\m\n}_{ZZ}$
into the definitions of the $S,T,U$ parameters we obtain \eqns\eqref{ResultT}, 
\eqref{ResultS} and \eqref{ResultU}. Since the one-loop integrals are divergent,
\begin{subequations}
\label{divergencies}
\begin{align}
Q^{\rm div}(q^{2},m^2_1,m^2_2)&=\epsilon^{-1}(m_{1}^{2}/2+m_{2}^{2}/2-q^{2}/3)\,,\\
B^{\rm div}_0(q^{2},m^2_1,m^2_2)&=\epsilon^{-1}\,,
\end{align}
\end{subequations}
each of the terms in these expressions is divergent as well. However, their combinations 
in \eqns\eqref{ResultT}, \eqref{ResultS} and \eqref{ResultU} are finite. This can be checked 
explicitly by using the unitarity of the full mixing matrix $\mathbf{U}$ as well as the 
relation $(m_L)_{\alpha\beta}=\sum_{i=1}^{3+n}\mathbf{U}_{\a i}m_{i}\mathbf{U}_{i\b}^{T}=0$,
where $\alpha,\beta \in \{e,\m,\t\}$. The latter 
reflects the fact that for a type-I see-saw the upper-left corner of the mass matrix is zero 
in the flavor basis. Since the divergences and the scale $\mu$ in the finite parts of the loop 
integrals always appear in the same combination, $\epsilon^{-1}+\ln \mu^2$, the cancellation 
of the divergences implies that the $\mu$-dependence drops out as well. In other words, the 
$S,T,U$ parameters depend only on physical quantities like the couplings, masses and momentum 
transfer. This is a consequence of the fact that they are defined in terms of the on-shell 
renormalized self-energies, see \eqn\eqref{STUdefinition}.

\section{\label{TwoCompNotation}Two-component notation}

To evaluate the contribution of the Majorana fermions to the self-energies  it is 
convenient to use the two-component spinor notation. In terms of the two-component 
spinors the four-com\-po\-nent Dirac spinor and its Dirac-conjugate read \cite{Plumacher:1998ex}: 
\begin{align}
\label{PsiDirac}
\Psi_D=\left(
\begin{array}{c}
\chi_\beta\\
\bar{\xi}^{\dot \beta}
\end{array}
\right)\,,\quad 
\bar \Psi_D=(\xi^\beta,\bar \chi_{\dot \beta})\,,
\end{align}
with $\xi=e_L$ and $\bar\xi=e_R$ for the charged leptons.
For Majorana fermions the two spinors in \eqref{PsiDirac} are not independent, $\xi=\chi$,
and therefore the number of degrees of freedom is reduced from four to two:
\begin{align}
\label{PsiMaj}
\Psi_M=\left(
\begin{array}{c}
\chi_\beta\\
\bar{\chi}^{\dot \beta}
\end{array}
\right)\,,\quad 
\bar \Psi_M=(\chi^\beta,\bar \chi_{\dot \beta})\,,
\end{align}
with $\chi=\nu$ for the Majorana neutrino. The Dirac matrices can be written in the form
\begin{align}
\gamma^\mu=\left(
\begin{array}{cc}
0 & \sigma^\mu\\
\bar\sigma^\mu & 0
\end{array}
\right)\,,\quad 
\gamma^5=\left(
\begin{array}{cc}
-\mathds{1} & 0\\
0 & \mathds{1}
\end{array}
\right)\,,
\end{align}
where $\bar \sigma^0=\sigma^0=\mathds{1}$ and $\bar \sigma^\mu=-\sigma^\mu$
with $\sigma^\mu$ being the Pauli matrices for $\mu=1,2,3$. Using this 
representation and the known formula for traces of the Dirac matrices we find:
\begin{subequations}
\begin{align}
\tr[\g^{\m}\g^{\n}]&=\tr[\sigma^{\m}\bar{\sigma}^{\n}]+\tr[\bar{\sigma}^{\m}\sigma^{\n}]\,,\\
\tr[\g^{\m}\g^{\n}\g^{\r}\g^{\l}]&=\tr[\sigma^{\m}\bar{\sigma}^{\n}\sigma^{\r}\bar{\sigma}^{\l}]+
\tr[\bar{\sigma}^{\m}\sigma^{\n}\bar{\sigma}^{\r}\sigma^{\l}]\,.
\end{align}
\end{subequations}
Using \eqref{PsiMaj} and the known form of the Feynman propagator of a Dirac field we 
can now infer the form of the propagator of a Majorana field:
\begin{align}
\label{MajPropagator}
\langle & T \Psi_M(x) \Psi_M(y) \rangle  =
\left(\hspace{-1mm}
\begin{array}{cc}
\langle T \chi_\beta(x) \chi^\gamma(y) \rangle &  \langle T \chi_\beta(x) \bar\chi_{\dot\gamma}(y) \rangle \\
\langle T \bar \chi^{\dot \beta}(x) \chi^\gamma(y) \rangle & \langle T \bar \chi^{\dot \beta}(x) \bar \chi_{\dot\gamma}(y) \rangle
\end{array}
\hspace{-1mm}\right)
\nonumber\\
& = i \int \frac{d^4 p}{(2\pi)^4}\frac{e^{-ip(x-y)}}{p^2-m^2+i\epsilon}
\left(
\begin{array}{cc}
m\, \delta^{\gamma}_{\b} & p_\mu \sigma^\mu_{\beta\dot\gamma} \\
p_\mu \bar\sigma^{\mu,\dot\beta\gamma} & m\, \delta^{\dot \beta}_{\dot\gamma}
\end{array}
\right)\,.
\end{align}
The diagonal components of \eqref{MajPropagator} describe contractions of the field with 
itself and reflect the Majorana nature of the field.

To use the above formulae we need to rewrite Lagrangian \eqref{gaugelagrangian} in terms 
of the two-component spinors:
\begin{align}
\label{gaugelagrtwocomp}
&\mathscr{L}=-\frac{e}{2c_ws_w}Z_\mu{\textstyle\sum_{i,j=1}^{3+n}}{\textstyle\sum}_\alpha
\bar{\nu}_{i,\dot\beta} \mathbf{U}^\dagger_{i\alpha}\bar\sigma^{\mu,\dot\beta \beta} 
 \mathbf{U}_{\alpha j} \nu_{j,\beta}\nonumber\\
&-\frac{e}{\sqrt{2}s_w} W_\mu{\textstyle\sum_{i=1}^{3+n}}{\textstyle\sum}_\alpha\bar{\nu}_{i,\bar\beta}
\mathbf{U}^\dagger_{i\alpha}\bar\sigma^{\mu,\dot\beta\beta} e_{L\,\alpha,\beta}+{\rm h.c.}\,,
\end{align}
where $\alpha$ are the flavor and $\beta$, $\dot{\beta}$  the spinor indices. As can be 
inferred from \eqref{gaugelagrtwocomp}, the contribution to the  $Z$ boson self-energy is 
proportional to
\begin{align}
\label{ContrtoZ}
\bigl\langle T\bigl[ \bar\nu_{\dot{\beta}}(x)\bar{\sigma}^{\m,\dot{\beta}\gamma}\nu_{\gamma}(x)
\bar\nu_{\dot{\rho}}(y)\bar{\sigma}^{\nu,\dot{\rho}\eta}\nu_{\eta}(y)\bigr] \bigr\rangle\,,
\end{align}
where we have suppressed the generation indices to shorten the notation. Now we need to 
use Wick's theorem and find all possible contractions. From \eqref{MajPropagator} it 
follows that for Majorana fermions there are two possibilities. The first is the contraction
of the field with its conjugate. This gives rise to
\begin{align}
\label{kineticcontr}
& - \langle T  \nu_{\eta}(y)\bar{\nu}_{\dot{\beta}}(x) \rangle 
\bar{\sigma}^{\m,\dot{\a}\b}
\langle T \nu_{\gamma}(x)\bar{\nu}_{\dot{\rho}}(y) \rangle
\bar{\sigma}^{\n,\dot{\g}\v}   \nonumber \\
& \propto -\tr[\,p\cdot\sigma\bar{\sigma}^{\m}k\cdot\sigma\bar{\sigma}^{\n}]=
-{\textstyle\frac12}\tr[\,\slashed{p}{\g}^{\m}\slashed{k}\g^{\n}]\,.
\end{align}
The second contraction possible only for Majorana fermions is the contraction of 
the field with itself. It gives rise to 
\begin{align}
\label{masscontr}
&- \langle T \bar\nu^{\dot{\beta}}(x)\bar{\nu}_{\dot{\rho}}(y) \rangle
{\sigma}^{\m}_{\gamma\dot{\beta}}
\langle T {\nu}_{\eta}(y)\nu^{\gamma}(x)\rangle
\bar{\sigma}^{\n,\dot{\rho}\eta}   \nonumber \\
& \propto -m_i m_j \tr[\sigma^{\m}\bar{\sigma}^{\n}]=-{\textstyle\frac12}m_i m_j \tr[\g^{\m}\g^{\n}]\,,
\end{align}
where we have used the identity 
\begin{align}
\bar\chi_{\dot\beta}\bar\sigma^{\mu,\dot \beta\gamma}\chi_{\gamma}=
-\chi^{\gamma}\sigma^{\mu}_{\gamma\dot \beta}\,\bar\chi^{\dot\beta}\,.
\end{align}
Collecting the two contributions and taking the traces of the Dirac matrices we find 
that \eqref{ContrtoZ} is proportional to:
\begin{align}
-2\left(\,p^{\m}k{}^{\n}+p^{\n}k-(pk)\, g^{\m\n}+m_i m_jg^{\m\n}\right).
\end{align}
The contribution to the self-energy of the $W$ boson is proportional to
\begin{align}
\label{ContrtoW}
\bigl\langle T\bigl[ \bar\nu_{\dot{\beta}}(x)\bar{\sigma}^{\m,\dot{\beta}\gamma}e_{L \gamma}(x)
\bar e_{L \dot{\rho}}(y)\bar{\sigma}^{\nu,\dot{\rho}\eta}\nu_{\eta}(y)\bigr] \bigr\rangle\,.
\end{align}
Because of the Dirac nature of the charged leptons only one, namely the `kinetic' contraction is possible
in this case. It results in an expression identical to \eqref{kineticcontr}. \\

\section{\label{MetropolisAlgorithm} Metropolis Algorithm}

In this work we minimize the $\chi^2$ function in a multidimensional parameter space. 
Due to the existence of strict constraints on the observables, the gradient minimization methods
are not efficient for finding global minima. Instead a statistical method is applied. 
In the first step we perform a random scan in the nine dimensional parameter space. The scan 
parameters are the three complex angles of the arbitrary orthogonal matrix in 
\eqn\eqref{active-sterile-mixing} and the three masses of the sterile neutrinos. 
The masses are chosen on a logarithmic scale to cover most efficiently a broad mass range. 

If a point  of the parameter space chosen in  this way satisfies the $\mu\rightarrow e\gamma$
and $0\nu\beta\beta$ constraints, a second step is performed. In the second step we calculate the full $\chi^2$ and perform a local
minimization. For this purpose we utilize the 
Metropolis algorithm which is usually taken to simulate phase transitions in the Ising model. First a fictive parameter (which we call temperature $\mathcal{T}$)
is introduced. Then parameters of the potential good-fit point are changed randomly by a small amount. 
The $\chi^2_\text{new}$ is computed for the new point and the Boltzmann function $B(\mathcal{T},\Delta{\chi^2})$ 
with the $\chi^2$ difference is evaluated:
\begin{align}
B(\mathcal{T}, \Delta{\chi^2}) = \exp{\left(-\frac{|\chi^2_\text{old}-\chi^2_\text{new}|}{\mathcal{T}}\right)} \,. 
\end{align} 
Furthermore, following the equal probability distribution, a random variable  $0<x<1$ is generated. 
If $\chi^2_\text{new}< \chi^2_\text{old}$  or $B(\mathcal{T}, \Delta{\chi^2})>x$ the new point is 
chosen as the new starting point. If $B(\mathcal{T}, \Delta{\chi^2})<x$ the new point is discarded. 
This process is repeated at a temperature $\mathcal{T}_1$ until a quasi-equilibrium is reached and 
no large changes in $\chi^2$ occur. Then the temperature is decreased to $\mathcal{T}_2<\mathcal{T}_1$
and  the process is repeated until the quasi-equilibrium is reached at the new temperature. This 
iterations continue until the effective temperature is zero. 

This method proves to be more efficient than a gradient method for finding global minima in 
the considered case since the system is highly constrained and thus the parameter space has a very non-trivial topology. 
Using the finite effective temperature approach we are able to scan a 
larger part of the parameter
space before the system settles in a minimum when the temperature drops to zero. Therefore, with a 
higher probability, the found minimum is the global one. 



\end{document}